\documentclass[aps,12pt,showpacs,nofootinbib]{revtex4}
\usepackage[english]{babel}
\usepackage{amsmath}
\usepackage{amssymb}
\usepackage{amsbsy}
\usepackage{amstext}
\usepackage{bm}
\usepackage{graphicx}
\usepackage{subfigure}
\usepackage[normalem]{ulem}
\usepackage{pst-node}

\newcommand{\be}{\begin{eqnarray}}
\newcommand{\ee}{\end{eqnarray}}
\newcommand{\nn}{\nonumber}
\newcommand{\bdm}{\begin{displaymath}}
\newcommand{\edm}{\end{displaymath}}
\newcommand{\ds}{\displaystyle}
\newcommand{\ba}{\begin{array}}
\newcommand{\ea}{\end{array}}
\newcommand{\pa}[1]{\left(#1\right)}
\newcommand{\paq}[1]{\left[#1\right]}

\newcommand{\K}{{\bf k}}
\newcommand{\Q}{{\bf q}}
\newcommand{\pp}{{\bf p}}
\newcommand{\X}{{\bf x}}

\newcommand{\ta}{{\tt a}}
\newcommand{\intkq}{\int_{\K\,,\Q}}
\newcommand{\eps}{\epsilon}

\begin{document}

\title{Near and far
  zone in two-body dynamics: \\
  an effective field theory perspective}

\author{Stefano Foffa$^{\rm 1}$ and Riccardo Sturani$^{\rm 2}$}

\affiliation{$(1)$ D\'epartement de Physique Th\'eorique and Center for Astroparticle Physics, Universit\'e de 
             Gen\`eve, CH-1211 Geneva, Switzerland\\
             $(2)$ International Institute of Physics, Universidade Federal do Rio Grande do Norte, Campus Universit\'ario, Lagoa Nova, Natal-RN 59078-970, Brazil}

\email{stefano.foffa@unige.ch, riccardo@iip.ufrn.br}

\begin{abstract}
  We revisit several aspects of the interaction of self-gravitating, slowly
  varying sources with their own emitted radiation within the context of
  post-Newtonian approximation to General Relativity.
  We discuss and clarify the choice of boundary conditions of Green's functions
  used to determine conservative potentials,
  and the interplay between the so-called near and far zones,
  as well as the relation between far zone ultra-violet divergences and emitted
  power.
  Both near and far zone contributions are required for the computation of the conservative dynamics.
  Within a field-theory approach we rederive far-zone self-energy
  processes, known as tail and
  memory effects, generalising the calculation of their divergent part to
  arbitrary order in the post-Newtonian expansion.
  \end{abstract}



\maketitle

\section{Introduction}

The recent detection of Gravitational Waves (GW) by compact binary
coalescences, see \cite{Abbott:2020niy} for the latest catalogue, by the large
interferometers LIGO \cite{TheLIGOScientific:2014jea} and Virgo
\cite{TheVirgo:2014hva}, marked the raising of Gravitational Astronomy onto the
stage of observational astronomy, with strong repercussion on fundamental
theoretical physics too.

To maximise the physics output of GW detection, observatory data are processed
via \emph{matched filtering} techniques \cite{Allen:2005fk}, which are
particularly sensitive to the phase of GWs. The GW-phase is in turn
determined by the dynamics of the sources, depending on the binary constituent's
astrophysical parameters and on the underlying gravitational law ruling their
motion.
Hence of paramount importance is the availability of accurate
template waveforms to correlate with the data,
the templates being determined by the solutions to the General Relativity two
body problem.

There exists today several approaches to solving the General Relativistic 2-body
problem, which all contributed to the waveform construction, even though
historically LIGO/Virgo waveform templates depended crucially on analytic
post-Newtonian (PN) approximation to General Relativity, see
\cite{Isoyama:2020lls} for a recent review.

A natural expansion parameter when dealing perturbatively with gravitational
interactions is Newton gravitational constant $G_N$, adopted by the
post-Minkowskian (PM) approximation of GR,
the Newtonian potential representing the 1PM approximation to two-body
scattering.
Methods borrowed from quantum field theory have been proven extremely useful
also for constructing a correspondence between the classical dynamics of bound
system and unbound scattering, thanks to which the PM analysis has been completed at
3PM order from first principle in \cite{Bern:2019nnu} and recently
extended to part of the 4PM order \cite{Bern:2021dqo}.
Note that the $n$PM order includes the $(n-1)$-loop order in Feynman diagram
expansion, i.e. next$^{n-1}$-to-leading order with respect to Newtonian
interaction. For another approach relating bound and unbound systems, see also \cite{Kalin:2020mvi}.

Another long-used approximation is the \emph{extreme mass ratio limit}, whose
parameter of expansion is the ratio of the masses of the binary constituents,
and whose results have currently reached next-to-leading order for bound-states,
see \cite{Pound:2021qin} for a recent review.

A novel, syncretic approach combining PN, PM and self-force results and
effective-one-body re-summation \cite{Damour:2019lcq} has led to a
systematisation of the centre-of-mass Hamiltonian, with partial knowledge of
the 5PN and 6PN dynamics, the missing sectors being the ones with higher
gravitational interactions ($G_N^5,G_N^6,G_N^7$)
\cite{Bini:2020nsb,Bini:2020uiq}.

In the post-Newtonian approach the parameter of expansion is $v^2$, being $v$
the relative velocity of the binary constituents, and since for a bound motion
$G_NM/r\sim v^2$, being $M$ the total mass of the binary system and $r$ the
distance between the binary constituents, all different terms (sub-sectors) of
the type $G_N^{1+n-j}v^{2j}$, for $0\leq j\leq n$, correspond to the same
$n$-PN order.

The PN approximation provides a framework to efficiently tackle the
2-body problem in GR which has now reached 5PN accuracy \cite{Foffa:2019rdf,Foffa:2019hrb,Blumlein:2019zku,Foffa:2019eeb,Foffa:2020nqe,Blumlein:2020pyo}, or next$^5$-to-leading order with respect to
Newtonian interaction, obtained within the approach named
Non Relativistic General Relativity (NRGR), pioneered by
\cite{Goldberger:2004jt}, see also \cite{Blumlein:2021txj} for 6PN result up to
$O(G_N^4)$ order included.

Crucial for the success of the PN program to reach this unprecedentedly high
perturbative order is the splitting of the analysis into
a \emph{near} (to the source) zone and a \emph{far} (or radiative) one.
In the former region the dynamics depends exclusively on
the exchange of longitudinal, off-shell modes, where Fourier representation of Green's functions can be
expanded for $k_0\ll |\K|$ in powers of $(k_0/|\K|)^2$,
being $(k_0,\K)$ the
$4$-dimensional argument of a Green's function.
Physically this correspond to expanding relativistic retardation effects in
gravitational mode exchange in powers of $v$, expressing
retarded times in terms of derivative of binary constituent trajectories.
Loop-integrals of Taylor-expanded Green's functions are then drastically
simplified with respect to computations involving the full, exact Green's
function,
enabling to perform the loop integral almost straightforwardly until 4PN order,
where the stumbling block of a single precedently unknown master integral was
identified and remarkably solved in \cite{Foffa:2016rgu}, paving the way to the
5PN and 6PN results mentioned earlier, which involved no new master integrals \footnote{On the other hand, the still unexplored 6PN, $O(G_N^5,G_N^6,G_N^7)$
  sectors are expected to contain new master integrals.}.

In principle at any PN order the complete result includes also the contribution from the far zone,
where the use of the full, unexpanded Green's function involves more difficult master integrals than in the near zone.
Following the \emph{method of regions}, contributions from near
and far zone must be added with unrestricted virtual momenta integration to
obtain the correct effective dynamics.
Overcounting, i.e. adding contribution from the same region of integration of
internal line momenta from both the near and the far zone, is avoided in dimensional
regularisation \cite{1972NuPhB..44..189T}, as described in general context in
\cite{Manohar:2006nz,Jantzen:2011nz} and the demonstration of how it is
avoided in the specific setting of NRGR is the first main goal of the present work.
The great advantage of working within the PN framework is that the lowest
perturbative order affected by far zone dynamics is the 4PN one, thus pushing to high
orders processes whose evaluation involves hard-to-compute integrals .

The second main goal of this paper is to revisit the issue of
which are the appropriate boundary conditions of Green's functions used in
the far zone (spoiler: causal boundary conditions, with caveats explained in the text).

The interplay between near and for zone computations, with the occurrence of
mutually compensating divergences, has been treated
in detail in \cite{Foffa:2019yfl} at 4PN level, the third main goal of
this paper is to generalise such discussion to arbitrary PN order at quadratic
level in Newton constant, showing that at this order in $G_N$
and at any order in the PN expansions, the so called \emph{tail} diagrams,
i. e. diagrams involving one conserved source and two dynamical ones, 
  contribute to both conservative at dissipative dynamics whereas \emph{memory} ones,
  i.e. diagrams involving dynamical sources only, do not contribute to
dissipation.

This paper is structured as follows: in Section \ref{sec:gen} we review the
role of boundary conditions in a generic field theory first, and then in the
specific case of NRGR. In Section \ref{sec:here}  we focus on NRGR diagrams with
one bulk triple interaction vertex, i.e. at $G_N^2$
or next-to-leading order self-energy diagrams, 
and we show explicitly that they split into contributions coming from the near
and from the far zones, discussing various examples.
Section \ref{sec:appl} contains the determination of the UV divergences
(which trace exactly power emission) at all PN orders with particular attention to 5PN.
Our concluding remarks are contained in Section ~\ref{sec:summ}.

\section{Feynman vs. causal boundary conditions}
\label{sec:gen}

\subsection{Boundary conditions for Feynman diagrams in field theory}
\label{ssec:bound_field}

We summarise in this section standard material about the appropriate choice of
Green's function boundary conditions to determine conservative dynamics in the
presence (or absence) of radiative modes.
Our goal is the application to the derivation of two-body effective Lagrangian
computed by integrating out gravitational modes exchanged between massive
sources via Feynman path integral.

Let us start by considering the toy model of a scalar field, whose generic Green's function
will be denoted by $G$, linearly coupled to two sources $J_{1,2}$.
At leading order in coupling one has the effective action schematically given
by
\be
\label{eq:eff1}
S_{eff}\sim \int dt_1dt_2 J_1(t_1)G(t_1-t_2)J_2(t_2)\,,
\ee
where spatial coordinates (and integration over) is understood.
Equations of motion can then be derived once a choice of boundary conditions
for the Green's function is made, choice that is
independent on the fundamental underlying theory, but which rather
depends on the specific physical setup under investigation.
In particular the potential generated at the position of particle 2 by particle
1 will be obtained convolving the \emph{retarded} Green's function with $J_2$.

Path integral methods have been originally developed within the context of
particle scattering, where differently from here the appropriate Green's function
is the Feynman one $G_F$, with particles ingoing for $t\to -\infty$ (outgoing for $t\to \infty$):
\be
\label{eq:Green_Feyn}
G_F(t,\X)\equiv \int_\K\frac{{\rm d}k_0}{2\pi} \frac{e^{-ik_0 t+i\K \cdot \X}}{k_0^2-\K^2+i{\tt a}}\,,
\ee
where we adopt the notation $\int_\K\equiv \int\frac{{\rm d}^dk}{(2\pi)^d}$ and
$'{\tt a}'$ stands for an arbitrary small positive number.\footnote{We did not adopt
  the $\epsilon$ character commonly used in literature, as we prefer
  to keep it to denote $d-3$, being $d$ the number of space dimensions.}
Note that in general $G_F$ is even under
$t\leftrightarrow -t$, $\X\leftrightarrow -\X$ exchange,
and it is complex in direct space as its Fourier transform,
which can be straightforwardly read from eq.~(\ref{eq:Green_Feyn}), shows that
$\tilde G_F(\omega,\K)\neq \tilde G_F^*(-\omega,-\K)$ and
$\tilde G_F(\omega,\K)\neq -\tilde G_F^*(-\omega,-\K)$.

Advanced and retarded and advanced Green's functions $G_{A,R}$ are defined by
\be
\label{eq:Green_causal}
G_{A,R}(t,\X)\equiv \int_\K\frac{{\rm d}k_0}{2\pi}
\frac{e^{-ik_0 t+i\K\cdot \X}}{\pa{k_0\mp i{\tt a}}^2-\K^2}
\stackrel{d=3}{=}-\frac 1{4\pi}\frac{\delta(t\pm |\X|)}{|\X|}\,,
\ee
where the $d\to 3$ limit has been written explicitly,
showing that $G_{A,R}$ are real functions in any dimensions as 
$\tilde G_A(\omega,\K)=\tilde G_R(-\omega,\K)$, and they fulfil
the relationship $G_A(t,\X)=G_R(-t,\X)$.
Introducing the Wightman functions $\Delta_{\pm}(t,\X)$ defined by
\be
\label{eq:Wght}
\Delta_\pm(t,\X)\equiv\int_\K \frac{e^{\mp ik_0t+i\K\cdot\X}}{2k}=
\int_\K\frac{{\rm d}k_0}{2\pi}\theta(\pm k_0)\delta(k_0^2-\K^2)e^{-ik_0t+i\K\cdot\X}\,,
\ee
one can find the relations\footnote{To derive (\ref{eq:Greens})
  the representation of the Heaviside
  theta function $\theta(t)=i\int_{-\infty}^\infty\frac{{\rm d}\omega}{2\pi}
  \frac{e^{i\omega t}}{\omega+i{\tt a}}$ is used.} \cite{Galley:2009px}
\be
\label{eq:Greens}
\ba{rcl}
\ds G_F(t,\X)&=&\ds-i\paq{\theta(t)\Delta_+(t,\X)+\theta(-t)\Delta_-(t,\X)}\,,\\
\ds G_R(t,\X)&=&\ds-i\theta(t)\paq{\Delta_+(t,\X)-\Delta_-(t,\X)}\,,\\
\ds G_A(t,\X)&=&\ds i\theta(-t)\paq{\Delta_+(t,\X)-\Delta_-(t,\X)}\,,
\ea
\ee
which enable to write the relation
\be
G_F(t,\X)=\frac 12\pa{G_A(t,\X)+G_R(t,\X)}-\frac i2\pa{\Delta_+(t,\X)+\Delta_-(t,\X)}\,.
\ee
The real part of the Feynman Green's function is given by the time-symmetric
combination $G_A+G_R$ and its imaginary part is given by the time-symmetric
combination of Wightman functions known as \emph{Hadamard} function.

As it will detailed in Section~\ref{sec:here}, the near zone
approximations within NRGR introduces a Taylor expansion of the Green's
functions in powers of $(k_0/|\K|)^2$,
making immaterial the $i{\tt a}$ part in the Green's function denominator:
in this approximation all Green's functions $G_F,G_A,G_R,(G_A+G_R)/2$ collapse to
the same quantity.
The Fourier space version of the Wightman functions have
support only on-shell, i.e. for $k_0=\pm |\K|$,
hence they vanish in the Taylor
expansions of the near zone which assumes $k_0\ll |\K|$.\footnote{Note that while
  ${\tilde \Delta}_{\pm}(k_0,\K)$ vanish off the light cone, this is not the case
  for the Fourier transform of $\theta(\pm t)\Delta_{\pm}(t,\X)$, as products in
  direct space translate to convolutions in Fourier space.}

On the other hand for modes that can go on-shell, which are involved in far
zone processes, the choice of Green's functions with Feynman or causal boundary
conditions leads to inequivalent results.
In this work we limit our analysis to processes with no outgoing radiation,
hence whenever a radiation mode is emitted and possibly scattered, it
will be eventually absorbed by the same system.
The processes of emission and absorption involve a time direction requiring the
use of advanced and retarded, or causal Green's functions.

In case of processes with \emph{only one} radiative mode involved,
considering that
internal line momentum integration is unrestricted over all real values $\omega$ of the
time-component of the momentum of the radiative mode,
the result of the amplitude is proportional to the time-symmetric combination
$G_A+G_R$, coinciding with the real part of the Feynman Green's function, see
Appendix \ref{app:tree}.
Note that while in Fourier space all Green's functions (Feynman, advanced and retarded) have non-vanishing imaginary part, in
  direct space causal Green's functions are real, as
  $\tilde G_{R,A}(\omega)=\tilde G^*_{R,A}(-\omega)$.
  
As a consequence no time-asymmetric effects (hence no dissipative effect)
are found, and what is obtained is the contribution of the exchange of
radiative modes to the conservative dynamics of the system.
Using the Feynman Green's function gives the correct real, time-symmetric part (which is vanishing for processes with one only, radiative mode involved)
and adds an imaginary part which is related via optical theorem
to the probability loss, which can be related to the emitted energy, as shown
in \cite{Foffa:2019eeb} for the gravitational case.

However it is possible to obtain radiation reaction effects exerted on the
source by its own emitted radiation, but this requires to adopt a different
Lagrangian than the one in (\ref{eq:eff1}). To take into account
non time-symmetric effects the sources $J_1$ and $J_2$ have to be treated
asymmetrically, as it is done in the \emph{in-in}
formalism and calculation requires doubling the degrees of freedom as performed
in the closed-time-path formalism \cite{Keldysh:1964ud}.
As we do not aim at computing radiation reaction forces here, we do not adopt
in-in formalism in this work and refer the interested reader to
\cite{Galley:2009px,Galley:2015kus}.

When two Green's functions of radiative modes are involved,
by using momentum conservation the process involves a product of the type
$\tilde G_R(\omega)\tilde G_A(-\omega)= \tilde G_R^2(\omega)$, with $\omega$ integrated over the entire real axis.

The difference between using causal Green's functions and two Feynman ones is
$\tilde G_F(\omega)^2-\frac 12\pa{\tilde G_R^2(\omega)+\tilde G_A^2(\omega)}\sim
-\tilde \Delta_+(\omega)\tilde \Delta_-(\omega)-i(\tilde G_A(\omega)+\tilde G_R(\omega))\tilde G_H(\omega)$,
where $\tilde G_H$ denotes the (Fourier transform of the)
Hadamard Green's function.
The real part of this difference vanishes because $\tilde \Delta_+(\omega)$ and
$\tilde \Delta_-(\omega)$ have no common support, the imaginary part 
is again related via the optical theorem to the probability loss, which also
in this case can be related to emitted energy.

For processes where three radiative modes are involved one encounters
products of the type
$\tilde G_R(\omega_1)\tilde G_R(\omega_2)\tilde G_A(-\omega_1-\omega_2)$ which
cannot be expressed purely in terms of Feynman Green's functions, as it is
also the case for a larger number of retarded/advanced Green's functions.

In the next section we apply these general considerations to the case of NRGR.

\subsection{Boundary conditions in NRGR}
\label{ssec:bound_NRGR}
The main application of NRGR is to build an effective Lagrangian for the two-body
problem in GR, by ``integrating out'' the gravitational degrees of freedom:
\begin{equation}
\label{eq:seff}
  S_{eff}\paq{x_a}=\int Dg_{\mu\nu} \exp\{i\pa{S_{EH+GF}\paq{g_{\mu\nu}}+ iS_{pp}\paq{x_a,g_{\mu\nu}}}\}\,,
\end{equation}
with
\be
\label{az_EH}
S_{EH+GF}&=& 2 \Lambda^2\int {\rm d}^{d+1}x\sqrt{-g}\paq{R(g)-\frac12\Gamma^\mu\Gamma_\mu}\,,\\
S_{pp}&=&-\sum_{a=1,2} m_i\int {\rm d}\tau_i = \sum_{a=1,2} m_i\int \sqrt{-g_{\mu\nu}(x^\mu_a) {\rm d}x_a^\mu {\rm d}x_a^\nu}\,,
\ee
where sources are described by non-dynamical worldline terms, for more details on our notations and conventions, see \cite{Foffa:2011ub}.
The functional integral in (\ref{eq:seff}) is evaluated pertubatively in the
$d+1$-dimensional Newton's constant $G\equiv \pa{32\pi\Lambda^2}^{-1}$ and expanded in the small
velocity limit:
this reduces the problem to the computation of a set of Feynman diagrams with explicit PN power counting.

We reiterate that for the purpose of obtaining effective Lagrangian terms
contributing to the conservative energy, causal boundary conditions for the
Green's functions should be adopted.

According to the standard PN construction,
velocity expansion implies that Green's functions for longitudinal
(i.e. non-radiative) modes are traded for a Taylor series in the near zone
\be
\label{eq:TayN}
\frac 1{\K^2-k_0^2}\longrightarrow N(k)\equiv \frac 1{\K^2}\sum_{n\geq 0}\pa{\frac{k_0^2}{\K^2}}^n\,,
\ee
and it will be useful to classify Feynman diagrams in the NRGR theory according to
how many full (i.e. non-expanded) Green's functions they contain.
It is understood that Taylor expansion (\ref{eq:TayN}) is valid only
  within its convergence radius, i.e. $k_0<|\K|$, and one has to take care
  when integrating Taylor-expanded Green's functions like in eq.~(\ref{eq:TayN}) over unrestricted region of momentum.

\begin{figure}
\includegraphics[width=.9\linewidth]{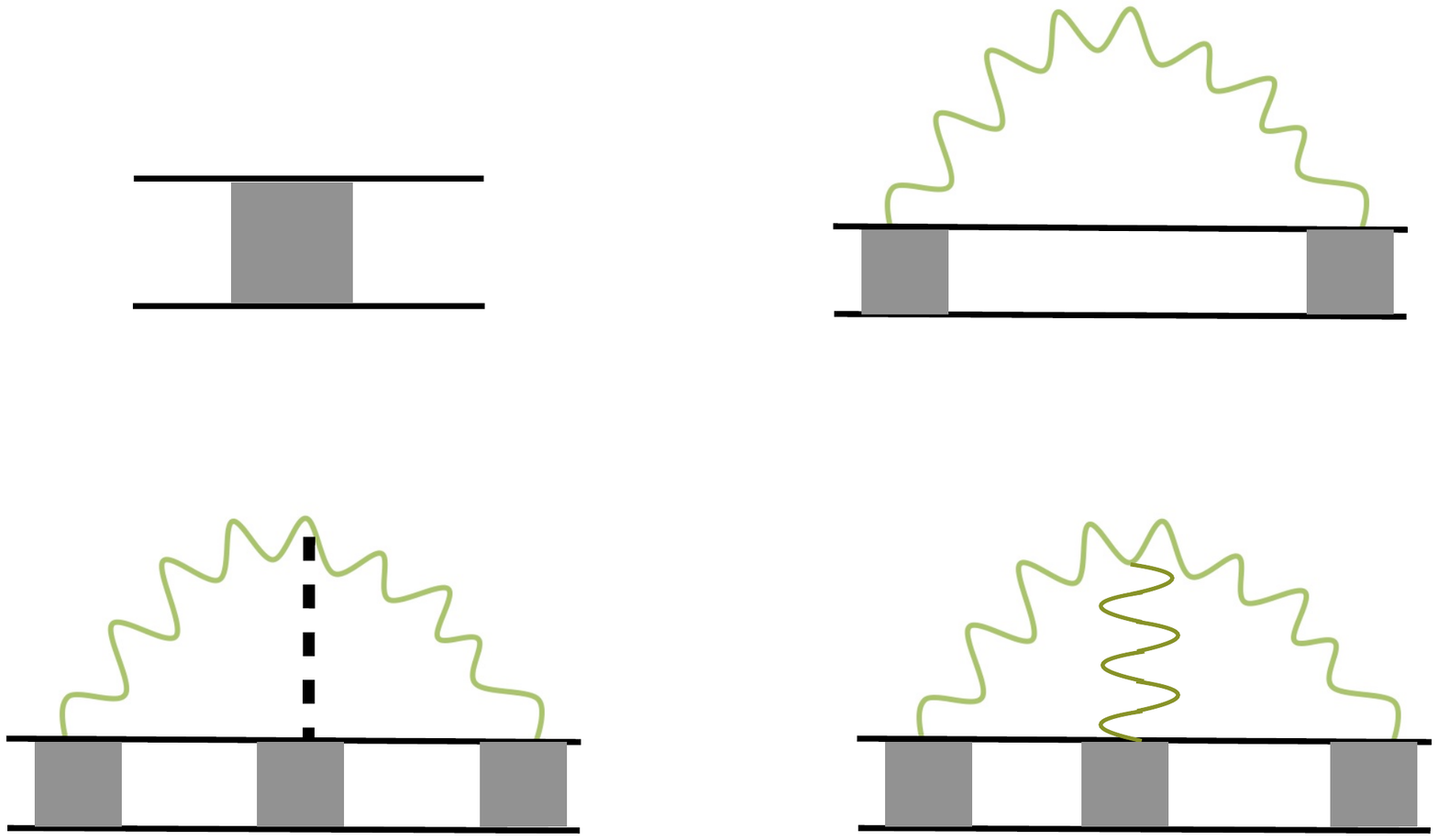}
\caption{Diagrams containing up to 3 full, non expanded, Green's functions.
  The horizontal continuous lines are black holes worldlines, the grey square
  represent any connected combination of Taylor-expanded (near zone)
  Green's functions.
  The dashed line in the bottom left diagram represents a Green's function with
  zero time-component momentum, i.e. representing a longitudinal, non-radiative mode sourced by a conserved charge.}
\label{fig:Mink0123}
\end{figure}

In the simplest case, corresponding to the upper left part of Figure \ref{fig:Mink0123}, all Green's functions have been Taylor-expanded: this is what happens
in near-zone diagrams and in this case the choice of boundary conditions is
irrelevant because none of the Green's functions is on-shell, as discussed in
Subsection \ref{ssec:bound_field}.

In diagrams containing just one exact (i.e.~not expanded as in (\ref{eq:TayN}))
Green's function, see upper right part of Figure \ref{fig:Mink0123}, as
discussed in previous Subsection the use of
Feynman Green's function gives the correct contribution to the
time-symmetric part of the effective action, which is vanishing \cite{Galley:2009px},
and the power emission can be derived from the imaginary part of the diagram.
The conservative part of equations of motion are not affected by this kind of
diagram (far-zone leading order self-energy), which however affect
their dissipative part through radiation reaction.
To capture dissipative effects one has evaluate such diagram in the in-in
formalism or closed-time-path formalism, which boils down to: (i) use of retarded
boundary conditions on the non-expanded Green's function, and (ii) doubling of
field and source variables to be able to derive dissipative effects in the equations of
motion directly from a variational principle.

The class of diagrams represented in the bottom left of Figure \ref{fig:Mink0123}
is called \emph{tail} \cite{Blanchet:1987wq}, and it is particularly
interesting because it is the lowest order diagram
containing exact Green's functions and yet giving a non-vanishing contribution
to the conservative equations of motion.
The wavy lines represent radiative modes sourced by dynamical multipoles, while
the dashed line represents a Green's function attached to a conserved charge,
hence its (Fourier transform of) Green's function has \emph{exactly zero} time
component momentum.
The real part of the diagram does not vanish, giving a finite contribution to the
4PN conservative 2-body dynamics (first derived in \cite{Foffa:2011np}),
which is necessary for the correct determination of conservative dynamics at
this PN order \cite{Bernard:2017bvn, Foffa:2019yfl}.
This diagram contains two radiative modes: by using Feynman boundary conditions
as in \cite{Foffa:2011np,Foffa:2019eeb} one obtains the correct conservative
dynamics, with the addition of the imaginary part related to the energy flux,
whereas causal boundary conditions would give the same result as
the Feynman Green's function for the real part, as explained in the previous
Subsection.

The bottom right process  in Figure \ref{fig:Mink0123} describes the memory
effect (or \emph{non-linear} memory-effect) \cite{Christodoulou:1991cr}, first
identified in the two-body problem in \cite{Blanchet:1992br}, where radiation is
scattered by radiation,
differently from the tail effect where radiation is scattered by the static
potential.
In the memory-type diagram two causal Green's functions must have the same
orientation from the source to the bulk vertex (retarded or advanced) while the
third one must have the opposite one (advanced or retarded).\footnote{
  The doubling of degrees of freedom required by the closed-time-path, or in-in
  formalism provides the configuration to recover the dynamics with
  non time-symmetric radiation reaction effects, see Appendix \ref{app:Keldysh}.}

A detailed calculation involving the contribution of these kind of diagrams
will be performed in Section \ref{sec:appl}.

\section{Near-Far zone interplay}
\label{sec:here}
When analysing 2-body dynamics within the PN approximation to General Relativity
it is customary to split the problem into a near and a far-zone.
In the former, longitudinal modes are exchanged between binary constituents,
which are described as fundamental object with standard coupling to gravity
and Green's functions are expanded as in eq.~(\ref{eq:TayN}).\footnote{Finite size effects affect dynamics only beyond 5PN order \cite{Damour:2009vw,Binnington:2009bb,Kol:2011vg}.}
In the latter both longitudinal and radiative modes are considered, with
exact Green's functions, and the binary systems is described as a composite
object in a multipolar expansion.

In this Section we show that addition of near and far zone contributions, with
unrestricted internal line momenta integration, returns the correct answer for
2-body dynamics at next-to-leading order in self-interactions
of radiative modes, paving the way for demonstration at higher $G$-interaction orders.
It will be shown how a single diagram in the exact theory gives rise to
different sub-processes spanning over all the four categories displayed in
Figure \ref{fig:Mink0123} when the domain of internal momentum integrals is split into
qualitatively distinct kinematical regions.

\subsection{Leading order}

To start with the simplest case, let us consider a process without gravitational
self-interaction in the bulk. 
In the full theory, its contribution to the effective action can be written as
\be
\label{eq:1prop}
      {\cal S}_1&=&\frac 1{\Lambda^2}\sum_{a,b=1}^2\ds\int_{t_{a,b}}\int_{k,p}
      C_a(t_a){\rm e}^{ik_0t_a-i\K\cdot \X_a(t_a)}C_b(t_b){\rm e}^{ip_0t_b-i\pp\cdot \X_b(t_b)}
      \frac 12\pa{\frac 1{R(k)}+\frac 1{A(k)}}\delta(k+p)
\ee
where $C_a(t_a)$ is the fundamental particle-gravity interaction vertex,
$\int_{t_{a,b}}$ is shorthand for $\int {\rm d}t_a {\rm d}t_b$,
$\int_k$ stands for the $d+1$-dimensional integral
$\int\frac{{\rm d}^{d+1}k}{(2\pi)^{d+1}}$, and
\be
\ba{rcl}
R(k)&\equiv&\ds \K^2-(k_0+i\ta)^2\,,\\
A(k)&\equiv&\ds \K^2-(k_0-i\ta)^2\,,
\ea
\ee
are the inverse causal Green's functions and note that (\ref{eq:1prop})
involves both $1\leftrightarrow 2$ interactions as well as self-interactions.

A standard, though non-trivial, result is that the integral over internal
momentum $k$ of the exact Green's function can be rewritten as a sum of the near zone approximation $N(k)$ of the Green's function, see eq.~(\ref{eq:TayN}),
plus the exact Green's function multiplying multipole-expanded source terms
obtained from the original ones by expanding the exponentials for
$|\K|r_a<1$, $r_a\equiv |\X_a|$:
\be
e^{ik_0t_a-i\K\cdot\X_a(t_a)}\stackrel{|\K|r_a<1}{=}e^{ik_0t_a}
\sum_{r\geq 0}\frac{\pa{-i\K\cdot \X_a(t_a)}^r}{r!}\,.
\ee
Eq.~(\ref{eq:1prop}) can then be re-written as
\be
\label{eq:1propNF}
\ba{rcl}
\ds{\cal S}_{1}&=&\ds\frac 1{\Lambda^2}
   \sum_{a,b=1}^2\int_{t_a,t_b}\int_k C_aC_b\Big[
     e^{ik_0t_{ab}-i\K\cdot \X_{ab}}N(k)+\\
&&\ds\qquad\left. \sum_{r,s\geq 0}\pa{\frac{\pa{-i\K\cdot\X_a}^r}{r!}}\pa{\frac{\pa{i\K\cdot\X_b}^s}{s!}}\frac 12\pa{\frac 1{R(k)}+\frac 1{A(k)}}\right]\,,
\ea
\ee
where $\X_a\equiv \X_a(t_a)$, $\X_{ab}\equiv \X_a-\X_b$, $C_a\equiv C(t_a)$, which
is exactly the sum of a near zone contribution (with the Green's function expanded)
and a far zone one in which the source has been expanded in multipoles.
Details of the derivation are reported in Appendix \ref{app:tree}.
Note that the contribution of the second line in (\ref{eq:1propNF}) is
purely imaginary, see eq.~(\ref{eq:fund}).

\subsection{Hereditary processes in NRGR}
\label{ssec:hered}
In the full theory, at next-to-leading order in gravitational coupling,
one has the contribution to the effective action from the process
involving one bulk triple interaction, endowed with a generic (and symmetrized)
bulk vertex $V$ at spacetime point $y$, which can be written as
\be
\label{eq:3props}
\ba{rcl}
{\cal S}_3&=&\ds\frac1{\Lambda^4}\int_{t_{a,b,c}}\int_{y}\int_{k,q,p}\frac{C_a{\rm e}^{-ik.\pa{x_a-y}}}{R(k)}\frac{C_b{\rm e}^{-iq.\pa{x_b-y}}}{R(q)}\frac{C_c{\rm e}^{-ip.\pa{x_c-y}}}{A(p)}V(k,q,p)\\
&=&\ds \frac 1{\Lambda^4}\int_{k,q,p}\delta^{(d+1)}\pa{k+q+p} I^{(R)}_a(k) I^{(R)}_b(q) I^{(A)}_c(p)V(k,q,p)\,,
\ea
\ee
where
\be
\ba{rcl}
\ds I^{(R)}_b(k)&\equiv&\ds\int_{t_b}{\rm e}^{ik_0 t_b}\frac{C_b{\rm e}^{-i\K\cdot\X_b}}{R(k)}\,,\\
\ds I^{(A)}_b(k)&\equiv&\ds\int_{t_b}{\rm e}^{ik_0 t_b}\frac{C_b{\rm e}^{-i\K\cdot\X_b}}{A(k)}\,,
\ea
\ee
and analogously
for $q,p$ and $a,c$. Lorentz indices eventually carried by $C$ and $V$ terms
are understood.

If all Green's functions are attached to the same particle,
the high momentum behaviour of the integrals is worsened,
and it can acquire UV divergences, which is the only
kind of divergence present in this kind of diagrams.
However in that case the result would involve only single particle variables,
hence it would not contribute to the two-body potential.

Following the causality considerations discussed in the previous Section, we
have to select two retarded and one advanced Green's functions, and since the
integration domain includes both positive and negative frequencies for each
of the internal line momenta, this includes also the other possible case (two
advanced and one retarded).

Starting from the full (post-Minkowskian) integral (\ref{eq:3props}) we
now split each spatial momentum integral into a \emph{hard} region, where the
condition $\{|\K|,|\Q|,|\pp|\}\geq\kappa$ is satisfied,
and a \emph{soft} region where $\{|\K|,|\Q|,|\pp|\}\leq\kappa$.
The conveniently chosen momentum scale $\kappa$ separates
the near-zone inverse length scale $r^{-1}$ from the far-zone frequency
scale $\Omega$, i.e. $r^{-1}\gg\kappa\gg\Omega$, where
$\Omega$ can be considered as the upper bound of the support of the
temporal component of the momentum $I^{(R,A)}_a$ depend on
\footnote{Strictly speaking, such quantities do not have a bounded support, even
  in the case of quasi-circular motion; however higher harmonics are more and more
  suppressed in the post-Newtonian setting, so being cavalier on this point does
  not spoil our proof as long as the orbital and the radiation scales are well
  separated.}.

In the case of a binary system it is straightforward to identify $r$ with the binary
constituent separation and $\Omega$ with $r/v$, being $v$ their relative
velocity.
Separation of scales and momentum conservation at the bulk triple vertex imply
that there can be none, one or three soft momenta in a physical process,
in no case exactly two of the momenta among $\K,\Q,\pp$ can be soft. 
For each term $I_a(k)$ we find useful to introduce a low velocity expansion
$N_a(k)$, a multipole expansion $F^{(R,A)}_a(k)$ and a double expansion $D_a(k)$
defined by:
\be
\label{eq:vexps}
\ba{rcl}
\ds N_a(k)&\equiv&\ds\int_{t_a}{\rm e}^{ik_0 t_a}\frac{C_a{\rm e}^{-i\K.\X_a}}{\K^2}\sum_{n\geq0}\pa{\frac{k_0^2}{\K^2}}^n\,,\\
\ds F^{(R,A)}_a(k)&=&\ds\int_{t_a}{\rm e}^{ik_0 t_a}\frac{C_a}{R[A](k)}\sum_{r\geq 0} \frac{\pa{-i\K.\X_a}^r}{r!}\,,\\
\ds D_a(k)&\equiv&\ds\int_{t_a}{\rm e}^{ik_0 t_a}\frac{C_a}{\K^2}\sum_{n\,,r\geq0}\pa{\frac{k_0^2}{\K^2}}^n \frac{\pa{-i\K.\X_a}^r}{r!}\,.
\ea
\ee

In terms of the Taylor-expanded terms (\ref{eq:vexps}), eq.(\ref{eq:3props})
can be trivially rewritten as
\be
\label{eq:full_nf}
\ba{rcl}
{\cal S}_3&=&\ds \frac 1{\Lambda^4}\int_{k,q,p}\delta^{(d+1)}\pa{k+q+p}V(k,q,p)\\
&&\times\left[N_a(k) N_b(q) N_c(p)+F^{(R)}_a(k) F_b^{(R)}(q)F^{(A)}_c(p)+S_{a,b,c}(k,q,p)\right]\,,
\ea
\ee
with $S_{a,b,c}(k,q,p)\equiv I^{(R))}_a(k) I^{(R)}_b(q) I^{(A)}_c(p)-N_a(k) N_b(q) N_c(p)-F^{(R)}_a(k) F_b^{(R)}(q) F^{(A)}_c(p)$.
We are now going to show that the 
last term in (\ref{eq:full_nf}), $S_{a,b,c}$, has vanishing real part for $p=-k-q$, with the consequence that the full theory
contribution (\ref{eq:3props}) to the conservative dynamics is obtained by
adding a pure near to a pure far zone term only.

Using the trivial identity $\theta(|\K|-\kappa)+\theta(\kappa-|\K|)=1$,
integrals over internal line momenta can be broken into hard and soft regions.
Moreover observing that in the hard region $I^{(R,A)}_a=N_a$ and $F^{(R,A)}_a=D_a$,
irrespectively of the boundary conditions, in the soft region
$I^{(R,A)}_a=F^{(R,A)}_a$ and $N_a=D_a$,
and that conservation of momentum forbids the case of one three-momentum only
among $\K,\Q,\pp$ to be hard, one can write, see Appendix \ref{app:HHS} for details,
\be
\label{eq:rest_props}
\ba{l}
\ds \delta^{(d+1)}(k+p+q)S_{a,b,c}(k,q,p)=\delta^{(d+1)}(k+q+p)\Big[-D_a(k)D_b(q)D_c(p)\\
\ds\quad +\theta(|\K|-\kappa)\theta(|\Q|-\kappa)\theta(\kappa-|\pp|)
  \paq{N_a(k)N_b(q)-D_a(k)D_b(q)}\paq{F^{(A)}_c(p)-D_c(p)}+{\rm perms}\Big]\,.
\ea
\ee
The first term on the right hand side of eq.~(\ref{eq:rest_props}) vanishes in
dimensional regularisation once integrated over $\K,\Q$ because it is a
scale-less integral\footnote{Scale-less integrals are integrals of the type
  $\int_\K\pa{\K^2}^\alpha$ and vanish in dimensional regularisation for any $\alpha$.
  Note however that they contain mutually cancelling infra-red and ultra-violet divergences for $d+2\alpha=0$, see Appendix \ref{app:mi}.}.
  
As to the remaining terms, one observes that the integration domain can be extended to the whole $\K,\Q,\pp$ space (taking into account momentum conservation)
because the first square bracket vanishes when $\K$ and $\Q$ are in the soft
region, and  the factor in the second square brackets of eq.~(\ref{eq:rest_props})
vanishes when $\pp$ is in the hard region.
Using then that all terms involving at least one of $D_{a,b,c}$ vanish when
integrated over unrestricted momentum because they are scale-less integrals,
one is left with
\be
\label{eq:full_simple}
\ba{rcl}
{\cal S}_3&=&\ds \frac 1{\Lambda^4}\int_{k,q,p}\delta^{d+1}\pa{k+q+p}V(k,q,p)\\
&&\times\left[N_a(k) N_b(q) N_c(p)+F^{(R)}_a(k) F_b^{(R)}(q)F^{(A)}_c(p)+N_a(k)N_b(q)F^{(A)}_c(p)+{\rm perms}\right]\,.
\ea
\ee
The final result shows the full amplitude in eq.~(\ref{eq:3props}) can be decomposed
into a pure near zone ($NNN$) contribution\footnote{The name, somehow misleading because the integration domain is the whole momentum space,
  refers to the fact that all Green's functions have been replaced by their low velocity expansion.},
a far zone ($FFF$) one, and a ``mixed''  ($NNF$ and permutations) contribution
which contains only one far-zone Green's function, thus belonging to the category
depicted in the top right part of Figure \ref{fig:Mink0123}, with one of the two near zone parts (the grey squares)
involving the triple bulk interaction vertex.

As discussed in Subsection \ref{ssec:hered}, diagrams with only one far-zone
Green's functions do not contribute to the conservative dynamics as they are
purely imaginary, see Appendix \ref{app:HHS} for details.

Note that the analysis of this Subsection applies also to the tail case, i.e. bottom left diagram of
fig.~\ref{fig:Mink0123}, in which out of the three modes parameterised
by the Green's functions two are radiative and one is longitudinal.
In the tail case one has a simplification:
one of the Green's functions is evaluated at vanishing time-momentum $\omega$
(because it is sourced by a conserved quantity, whose Fourier transform has support only at $\omega=0$),
hence it collapses to the leading term of (\ref{eq:TayN}), allowing the result
to be written in terms of the real part of the product of Feynman Green's
functions as explained in the previous Section, property which is not shared by
processes involving three causal Green's functions.

\subsection{The $\phi^3$ case in detail}
\label{ssec:phicube}
To better elucidate how pure near and far zone diagrams can reconstruct the full theory,
we now give a concrete example of how the method of regions work in a specific,
simple process at $O(G^2)$.
We find useful to adopt the following parameterisation of metric \cite{Kol:2007bc}
\be
\label{met_nr}
g_{\mu\nu}=e^{2\phi/\Lambda}\pa{
\ba{cc}
-1 & A_j/\Lambda \\
A_i/\Lambda &\quad e^{- c_d\phi/\Lambda}\gamma_{ij}-
A_iA_j/\Lambda^2\\
\ea
}\,,
\ee
with $\gamma_{ij}=\delta_{ij}+\sigma_{ij}/\Lambda$ and $c_d=2\frac{(d-1)}{(d-2)}$,
gravitational modes exchanged between massive particles can then be decomposed
into $\phi,A,\sigma$ according to their polarisation.

To illustrate the decomposition of the integrations into regions of momenta
it is sufficient to consider just the leading order term in the PN-expansion
for each region and we take the simplest case of the three modes being
$\phi$-polarised.
In this case the vertices are $C_a=-m_a$
(which we will assume to be time-dependent to make the argument more general)
and $V(k,p,q)=\pa{k_0p_0+q_0p_0+k_0q_0}/(3\times 8c_d)$.
The treatment of the other polarisation channels is qualitatively the same,
with different expressions for both bulk and worldline vertices.

\subsubsection{Near zone}
The near zone computation results into the following:
\be
\label{eq:ir_uv}
\ba{rcl}
\ds{\cal S}_{\phi^3near}&=&\ds 1024\pi^2 G^2\sum_{a,b,c=1}^2\int_{k,q,p}\delta^{(d+1)}\pa{k+q+p}V(k,q,p)N_a(k) N_b(q) N_c(p)\\
&=&\ds -\frac{128}{c_d}\pi^2 G^2\sum_{a,b,c=1}^2\int_{k,q}\int_{t_{a,b,c}}{\rm e}^{i\pa{k_0t_{ac}+q_0t_{bc}}}{\rm e}^{-i\pa{\K\cdot\X_{ab}+\Q\cdot\X_{bc}}}\frac{m_a m_b m_c}{\K^2 \Q^2 \pa{\K+\Q}^2}k_0 q_0\\
&=&\ds\frac{128}{c_d}\pi^2 G^2\int_t\int_{\K,\Q}\paq{\frac{\dot{M}^2 M}{\K^2 \Q^2 \pa{\K+\Q}^2}{\rm e}^{-i\K.{\bf r}}+
\frac{\dot{m}^2_1 m_1+\dot{m}^2_2 m_2}{\K^2 \Q^2 \pa{\K+\Q}^2}\pa{1-{\rm e}^{-i\K\cdot{\bf r}}}}\\
&\simeq&\ds G_N^2\int_t\paq{\frac1{\epsilon_{IR}} \dot{M}^2 M
  -\frac1{\epsilon_{UV}}\pa{\dot{m}^2_1 m_1+\dot{m}^2_2 m_2}+O(\eps^0)}\,,
\ea
\ee
where we have used the standard integrals listed in app.~\ref{app:mi},
introduced the notation $M\equiv m_1+m_2$ and in the last line the limit
$\epsilon\equiv d- 3\to 0$ has been taken, which makes appear the $3+1$-dimensional Newton's
constant related to the one in generic $d+1$ dimensions via
$G_N=\pa{\mu\sqrt{e^{-\gamma}/(4\pi)}}^\epsilon G$.
Note that we have distinguished between ultra-violet (UV) and infra-red (IR)
divergences, see Appendix \ref{app:mi} and \cite{Foffa:2019yfl} for derivation.

This corresponds to a 2PN near zone diagram and it is vanishing because
$\dot{m}_{1,2}=0$, but it gives an off-shell non-vanishing contribution to
the conserved dynamics (all near zone diagrams belong exclusively to the
conservative sector) starting from 3PN \cite{Foffa:2011ub,Foffa:2012rn}.

\subsubsection{``Mixed'' zone}
The leading order of the ``mixed'' term of the type $NNF$ in eq.~(\ref{eq:full_simple}) factorises into a term depending on $\K$ and one depending on $\pp$,
see Appendix \ref{app:HHS} for details, and it can be written as follows:
\be
{\cal S}_{\phi^3mix}=
\frac{128}{3c_d}\pi^2 G^2\sum_{a,b,c}\int_{k_0,q_0}\pa{-i q_0}\pa{k_0^2+q_0^2+k_0q_0}\int_{t_{a,b,c}}{\rm e}^{i\pa{k_0t_{ac}+q_0t_{bc}}}m_a m_b m_c\int_{\K}
\frac{{\rm e}^{-i\K.\X_{ac}}}{\K^{4}}\,,
\ee
where the explicit  expression of $V(k,q,p)$ has been used and the integral over
$\Q$ performed as per eq.~(\ref{eq:fund}).
Using the standard integral (\ref{eq:masterF}) one obtains the smooth
limit for $d\to 3$
\be
\ba{rcl}
\ds{\cal S}_{\phi^3mix}&=&\ds-\frac 43 \pi i G_N^2 \sum_{a,b,c}\int_{k_0,q_0}q_0
\pa{k_0^2+q_0^2+k_0q_0}\int_{t_{a,b,c}}{\rm e}^{i\pa{k_0t_{ac}+q_0t_{bc}}}m_a m_b m_c r_{ac}\\
&=&\ds-\frac 43 \pi G_N^2\sum_{a,b,c}\int_t m_c r_{ac}
\pa{\ddot m_a\dot{m}_b+m_a\dddot m_b+\dot m_a \ddot m_b}
\ea
\ee
showing that this terms is purely imaginary in Fourier representation, or
equivalently odd under time reversal in direct space, as it represents an
$O(G)$ correction to the far piece of (\ref{eq:1propNF}).

As already pointed out in the general discussion, this term is not conservative because it contains an odd number of time derivatives.
It has to be considered as the classical dressing in the $\phi^2$ channel of
the simple self-energy diagram (where one of the $M$'s is corrected
by the Newtonian potential),
which would give the power emitted in monopole radiation, in our toy model
in which total energy is time dependent.

\subsubsection{Far zone}
The analogous process in the far zone (soft momentum region) gives the following
amplitude:
\be
\ba{rcl}
\ds   {\cal S}_{\phi^3far}&=&\ds1024 \pi^2 G^2\int_{k,q,p}\delta^{(d+1)}\pa{k+q+p}V(k,q,p)
   F_a^{(R)}(k) F_b^{(R)}(q)F^{(A)}_c(p)\\
   &=&\ds\frac{128}{3c_d}\pi^2 G^2\sum_{a,b,c}\int_{t_{a,b,c}}\int_{k,q}
   {\rm e}^{i\pa{k_0t_{ac}+q_0t_{bc}}}\frac{m_am_bm_c}{R(k)R(q) A(-k-q)}
   \pa{k_0^2+q_0^2+k_0q_0}\,.
   \ea
\ee
Remarkably the double $d$-dimensional integral over $\K,\Q$ included above
  has been computed in \cite{Davydychev:1992mt}, as it
  is involved in gauge-theory 2-loop self energy computations,
  see Appendix \ref{app:mi} for details. Its divergent part is
  independent on Green's function boundary conditions and it gives
  
  \be
\ba{rcl}
\ds{\cal S}_{\phi^3far}&\simeq&\ds
-\frac1{3\epsilon_{UV}} G_N^2\int_{k_0,q_0}\int_{t_{a,b,c}}{\rm e}^{i\pa{k_0t_{ac}+q_0t_{bc}}}M(t_a) M(t_b) M(t_c)\pa{k_0^2+q_0^2+k_0q_0}+O(\eps^0)\\
&=&\ds -\frac1{\epsilon_{UV}}G_N^2\int_{t}\dot{M}^2 M+O(\eps^0)\,,
\ea
\ee
which cancels the near zone IR divergence as discussed in \cite{Foffa:2019yfl}.

Note that in this case it is not difficult to perform the calculation
of the divergent part in the full theory, which involves the same master integral as the far zone
  computations
  \be
  \ba{rcl}
  \ds{\cal S}_{\phi^3full}&=&\ds\frac{128}{3c_d}\pi^2 G^2\sum_{a,b,c}\int_{t_{a,b,c}}
  \int_{k,q}{\rm e}^{i\pa{k_0t_{ac}+q_0t_{bc}}}{\rm e}^{-i\pa{\K\cdot x_{ab}+\Q\cdot\X_{bc}}}
  \frac{m_am_bm_c}{R(k)R(q) A(-k-q)}\pa{k_0^2+q_0^2+k_0q_0}\\
  &=&\ds-\frac1{\epsilon_{UV}}G_N^2\int_{t}\dot m_1^2 m_1+\dot m_2^2m_2+O(\eps^0)\,,
  \ea
  \ee
  with the UV divergence coming exclusively from diagrams in which Green's functions are attached to the same particles.

\subsubsection{Far zone, tail}
In the previous toy examples, we artificially allowed time dependence on $m_a$
to illustrate the basic features of the region decomposition.
In a physical setting, the lowest PN order non-vanishing contribution from
the $\phi^3$ channel is obtained by taking the second term
(i.e. the quadrupolar term) in the multipolar series defining $F^{(R,A)}$ in two
out of three far zone propagation insertions.
Using that $m_a$ is constant in time also modifies the
causality structure of the diagram, and eliminates the pole from one of the
Green's functions
\be
\label{eq:tailphi3}
\ba{rcl}
\ds{\cal S}^{(Q^2M)}_{\phi^3}&=&\ds-\frac{32\pi^2}{c_d} G^2\int_{k,q,p}\delta^{(d+1)}\pa{k+q+p}\int_{t_{a,b,c}}{\rm e}^{i\pa{k_0t_a+q_0t_b+p_0t_c}}\frac{k_0p_0+q_0p_0+k_0q_0}{R(k)A(q)\pa{\K+\Q}^2}\\
&&\ds\qquad\qquad\times\sum_{a,b,c}\paq{\pa{-i\K.\X_a}^2 \pa{-i\Q.\X_b}^2+2\pa{-i\K.\X_a}^2 \pa{-i\pp.\X_c}^2}m_a m_b m_c\\
&=&\ds\frac{32\pi^2}{c_d} G^2M\int_{k_0}k_0^2\tilde{Q}_{ij}(k_0)\tilde{Q}_{mn}(-k_0)\int_{\K,\Q}\frac{k^i k^j q^m q^n}{\paq{\K^2-\pa{k_0+i \tt{a}}^2}\paq{\Q^2-\pa{k_0+i \tt{a}}^2}\pa{\K+\Q}^2}
\ea
\ee

For the point of view of sub-process computation, this is just a specific
  case of the hereditary integral (\ref{eq:3props}), in which one of the $C_a$
  terms is time-independent, hence its Fourier transform
  $\tilde C_a(k_0)=C_a\delta(k_0)$.

Equation ~(\ref{eq:tailphi3}) represents the contribution from intermediate
$\phi$-polarised gravitational modes to the complete tail process,
whose complete result, adding all intermediate polarisations, can be found in
\cite{Foffa:2019eeb}.

\subsubsection{Far zone, memory}

Analogously to the tail case, by expanding at second order all three far zone
source insertions, one obtains the $\phi^3$ contribution to the quadrupole
memory
\be
\label{eq:QQQret}
\ba{rcl}
\ds{\cal S}^{(Q^3)}_{\phi^3}&=&\ds-\frac{16\pi^2}{c_d} G^2\int_{k,q,p}\delta^{(d+1)}\pa{k+q+p}\int_{t_{a,b,c}}{\rm e}^{i\pa{k_0t_a+q_0t_b+p_0t_c}}\frac{k_0p_0+q_0p_0+k_0q_0}{3R(k)R(q)A(p)}\\
&&\ds\qquad\qquad\times\sum_{a,b,c}\pa{-i\K.\X_a}^2 \pa{-i\Q.\X_b}^2\pa{-i\pp.\X_c}^2m_a m_b m_c\\
&=&\ds\frac{16\pi^2}{3c_d} G^2M\int_{k_0,q_0,p_0}\delta\pa{k_0+q_0+p_0}\tilde{Q}_{ij}(k_0)\tilde{Q}_{mn}(q_0)\tilde{Q}_{kl}(p_0)\pa{k_0p_0+q_0p_0+k_0q_0}\\
&&\ds\quad\quad\times\int_{\K,\Q,\pp}\delta^{(d)}\pa{\K+\Q+\pp}\frac{k^i k^j q^m q^np^k p^l}{\paq{\K^2-\pa{k_0+i \tt{a}}^2}\paq{\Q^2-\pa{q_0+i \tt{a}}^2}\paq{\pp^2-\pa{p_0-i \tt{a}}^2}}\,.
\ea
\ee
The total memory term is computed by adding analogous contributions from all other
polarisations. The integrals can be computed with the help of the formulae
reported in Appendix B of
\cite{Foffa:2019eeb}, individual contributions from specific
polarisations contain the UV divergent integral $I_m$, but the sum turns out to
be finite (more about this in the next Subsection), and the final result gives
\be
\label{eq:mem}
S^{Q^3}_{eff\,5PN}&=&\ds -\frac{G_N^2}{15}
\int {\rm d}t \left[\ddddot{Q}_{il}\ddddot {Q}_{jl}{Q}_{ij}+\frac4{7}\dddot{Q}_{il}\dddot {Q}_{jl}\ddot{Q}_{ij}\right]\,.
\ee

\section{Far zone poles at 5PN and beyond}
\label{sec:appl}

The generic process at next-to-leading order in $G_N$ can be schematically written as
\be
\label{eq:her_gen}
A_{her}=\intkq \frac{{\rm d}q_0}{2\pi}\frac{{\rm d}k_0}{2\pi}M^{(1)}_{i_1\ldots i_l}(q_0)
M^{(2)}_{j_1\ldots j_m}(k_0)M^{(3)}_{k_1\ldots i_n}(-k_0-q_0)
\frac{P^{i_1\ldots i_lj_1\ldots j_mk_1\ldots k_n}(\K,\Q,k_0,q_0)}{\cal D}\,,
\ee
with $M^{(i)}_{i_1\ldots i_n}$ being the (Fourier-transformed) generic multipole
moment and the factor ${\cal D}\equiv({\bf k}^2-k_0^2)({\bf q}^2-q_0^2)[({\bf k}+{\bf q})^2-(k_0+q_0)^2]$, with the $i{\tt a}$ prescription now understood.

A generic feature of the double $\K,\Q$ integral in tail processes is the
concurrent appearance of divergences in the real part and finite imaginary
terms, when Feynman Green's functions are used, reminding that in tail
processes one of the sources is conserved obliging the relative
propagator to have vanishing time-component momentum.
Observing that divergences can only appear in the second integration because
of momentum exponent mismatch (integration is over three-dimensional space,
integrand have even power of momentum), the pole obtained after second
space-momentum integration is of the type:
$$\frac{G^2_d}\epsilon(-k_0^2-i{\tt a})^\epsilon=
G_N^2\pa{\frac 1\epsilon+\log(k_0^2/\mu^2)-i\pi+O(\epsilon)}\,,$$
as first observed in \cite{Foffa:2019eeb}.
In the remaining of this section this fundamental property will be exploited
to relate far-zone UV poles from tail diagrams to flux emission coefficients
for generic multipoles, whose simple, generic analytic expression was first
derived in \cite{Thorne:1980ru}.

The memory integral, where all three sources are dynamical, is also
described by a term like eq.~(\ref{eq:her_gen}) but as it will be shown in the
next subsection such processes are finite and do not contribute to dissipative
dynamics, even if when broken in terms of different gravitational polarisations,
some individual contributions diverge, as they involve the master integral (\ref{eq:Imem}), but the combined physical result is finite.

\subsection{UV poles at 5PN}
\label{sec:gnsqall}
We are now going to extend here to 5PN order the computation of UV divergent part
of next-to-leading order in $G_N$ self-energy amplitudes of the type (\ref{eq:her_gen}), which have been computed up to 4PN order (with their finite part)
in \cite{Foffa:2019yfl}.

The far zone UV divergences cancel as per the \emph{zero bin mechanism}
\cite{Manohar:2006nz}
with IR divergences from the near zone, as shown explicitly in the simple
$\phi^3$ example of Subsection \ref{ssec:phicube}.
While the far zone has no IR divergences, the near zone also possesses UV
ones, which can be absorbed by local counter-terms fully determined at
4PN order in \cite{Foffa:2019yfl}.

At 5PN order however the near zone divergences are known only overall, i.e.
UV plus IR together, as they have been determined in harmonic gauge in
\cite{Blumlein:2020pyo} (with partial 6PN results recently appeared in \cite{Blumlein:2021txj}).
Moreover 5PN near zone counter-terms are presently known only up to $G^2$ order,
as they are inherited from the 4PN calculation, but eventual
additional 5PN ${\cal O}(G^3)$ ones may be required and we plan
to determine them in future work.
In the present section we determine the far zone UV poles, which will be matched
to near zone IR ones in future work.

The starting point is the fundamental coupling between the energy momentum tensor $T_{\mu\nu}$ and the gravitational field $h_{\mu\nu}$
\be
\label{eq:mult_bis}
\ds S_{pp}=\ds\frac12\int T^{\mu\nu}(x)\frac{h_{\mu\nu}(\vec{x},t)}\Lambda{\rm d}t {\rm d}^d x\,.
\ee
According to standard multipolar expansion, it can be expressed in terms
of metric ansatz (\ref{met_nr}) as
\be
\label{eq:multgen}
\ba{rcl}
\ds S_{pp}&=&\simeq \ds\frac{1}{\Lambda}\int {\rm d}t \sum_{n\geq 0}\frac{1}{n!}\left[{\cal O}^{N}_{\phi} \phi_{,N}+{\cal O}^{iN}_{A} A_{i,N}+\frac12 {\cal O}^{ij N}_{\sigma} \sigma_{ij,N} \right] \,,
\ea
\ee
where the collective index $N\equiv i_1\dots i_n$ has been introduced and
the generic multipoles ${\cal O^N}$ are defined by
\be
\label{eq:OpAs}
  \ba{rcl}
\ds\tilde{\cal O}^{N}_{\phi}(t)&\equiv&\ds \tilde{\cal O}^{i_1\dots i_n}_{\phi}(t)= \int_x T^{00}x^{i_1}\dots x^{i_n}\\
\ds {\cal O}^{i N}_{A}(t)&\equiv&\ds {\cal O}^{i i_1\dots i_n}_{A}(t)= \int_x T^{0i}x^{i_1}\dots x^{i_n}\\
\ds {\cal O}^{ij N}_{\sigma}(t)&\equiv&\ds {\cal O}^{ij i_1\dots i_n}_{\sigma}(t)= \int_xT^{ij}x^{i_1}\dots x^{i_n}\\
\ds {\cal O}^{N}_{\phi}(t)&\equiv&\ds  -\tilde{{\cal O}}^{kk N}_\sigma(t) - {\cal O}^{N}_\phi(t)\,.
\ea
\ee
This results can be written in terms of the following (non-traceless) momenta:
\begin{eqnarray}
\label{eq:momenta}
&& E(t)\equiv\int_x T^{00}\,,\quad G^i(t)\equiv\int_x T^{00}x^i\,,\quad Q^{ij}(t)\equiv\int_x T^{00}x^i x^j\,,\\
&&O^{ijk}(t)\equiv\int_x T^{00} x^i x^j x^k\,,\ H^{ijkl}(t)\equiv\int_x T^{00} x^i x^j x^k x^l\,,\ D^{ijklm}(t)\equiv\int_x T^{00} x^i x^j x^k x^lx^m\,,\\
&& P^i(t)\equiv\int_x T^{0i}\,,\quad M^{iN}(t)\equiv \int_x T^{0i} x^N\,,\quad {\cal T}^{ijN}(t)\equiv \int_x T^{ij} x^N\,.
\end{eqnarray}

The general formulae of the Appendix \ref{app:poles_her} allow us to fully
recover the 4PN results displayed in Appendix B of \cite{Foffa:2019yfl}, except
for equation (B.4) which had been mistyped; we report here the correct version (using the same notation as in \cite{Foffa:2019yfl})
\be
{\cal L}_{A\phi^2,\, \rm 4PN}^{\rm UV\,(far)}&=&\frac{G^2}{3\epsilon_{\rm UV}}\left[ 4 \ddot {\bf X}^i{\bf P}^i\left(\dot M+ T^{kk(1)}\right)+\frac{2}{5}\dot{M}_{ij}\left(2\ddot{\bf X}^i\ddot{\bf X}^j+\ddot{\bf X}^k\ddot{\bf X}^k\delta^{ij}\right)+\frac{2}{5}I_0^{ij(3)}\left(2{\bf P}^i\ddot{\bf X}^j+{\bf P}^k\ddot{\bf X}^k\delta^{ij}\right)\right]\,.\nn
\ee

The new 5PN result are reported for each polarisation channel:
\be
\epsilon_{UV}{\cal L}^{\phi^3}_{5PN}&=&\frac1{15} \ddot{E}  \dot{G}^i \dddot{O}^{ikk}-\frac16\dot{E}^2 \ddot{Q}-\frac1{30}  \dot{E}\dddot{Q}^{ij}\left[\ddot{Q}\delta_{ij}+2 \ddot{Q}_{ij}\right]
-\frac1{60} E\dot{E}  \ddot{\dddot{H}}^{jjkk}\nn\\
&&-E\paq{\frac1{420} \ddot{G}^k \dddot{\dddot{D}}^{iijjk}+\pa{\frac1{2520} \dddot{Q}^{ab} \ddot{\dddot{H}}^{ijkc}+\frac1{3780} \ddddot{O}^{abc} \ddddot{O}^{ijk}}\delta_{ijkabc}}\nn\\
&&+\left[\frac1{2520} \ddddot{H}^{ijkc} \pa{ 2\dddot{G}^{a} \dot{G}^{b}+\ddot{G}^{a} \ddot{G}^{b}}+\frac1{630}\dddot{O}^{ijk}\pa{\ddddot{Q}^{ab} \dot{G}^{c}+\dddot{Q}^{ab} \ddot{G}^{c}+\ddot{Q}^{ab} \dddot{G}^{c}}\right.\nn\\
&&\left.\quad\quad-\frac1{840}\dddot{Q}^{ij} \dddot{Q}^{kc} \ddot{Q}^{ab}\right]\delta_{ijkabc}\,,\\
\epsilon_{UV}{\cal L}^{\phi^2\sigma}_{5PN}&=&\frac23\dot{E}^2 {\cal T}^{kk}+\frac2{15}\dot{E}\paq{2\dddot{Q}{\cal T}^{kk}-\dddot{Q}^{ij}{\cal T}_{ij}+2\ddot{G}^i\pa{2\dot{\cal T}^{kki}-\dot{\cal T}^{ikk}}}\nn\\
&&\frac1{315}\ddddot{O}^{abi}\ddot{G}^j\left(7{\cal T}^{kk}\delta_{ijab}-{\cal T}^{kl}\delta_{ijabkl}\right)+\frac1{210}\ddot{G}^i\ddot{G}^j\left(7\ddot{\cal T}^{kkab}\delta_{ijab}-\ddot{\cal T}^{klab}\delta_{ijabkl}\right)\nn\\
&&+\frac1{105}\dddot{Q}^{ij}\ddot{G}^k\left(7\ddot{\cal T}^{aal}\delta_{ijkl}-\ddot{\cal T}^{abl}\delta_{ijklab}\right)+\frac1{420}\dddot{Q}^{ij}\dddot{Q}^{kl}\left(7\ddot{\cal T}^{aa}\delta_{ijkl}-\ddot{\cal T}^{ab}\delta_{ijklab}\right)\,,\\
\epsilon_{UV}{\cal L}^{\phi A^2}_{5PN}&=&\frac1{15} E\paq{\ddot{\dddot{M}}^{ikkll}\dot{P}^i+4\ddddot{M}^{ijkk}\ddot{M}^{ij}+\dddot{M}^{ijk} \dddot{M}^{ilm}\pa{2\delta_{jl}\delta_{km}+\delta_{jk}\delta_{lm}}}+\frac{1}{30}\ddddot{O}^{kkll}\dot{\vec{P}}^2\nn\\
&&+\frac{4}{15} \ddot{M}^{ij}\dot{P}^i \dddot{O}^{jkk}+\frac{4}{15} \ddddot{M}^{ijkk}\dot{P}^i \dot{G}^{j}+\frac{2}{15} \ddot{M}^{ij}\ddot{M}^{ik}\pa{\ddot{Q}\delta_{jk}+2\ddot{Q}_{jk}}
+\frac{4}{15} \dddot{M}^{ijk}\ddot{M}^{il}\dot{G}^m\delta_{jklm}\nn\\
&&+\frac{2}{15} \dddot{M}^{ijk}\dot{P}^{i}\pa{\ddot{Q}\delta_{jk}+2\ddot{Q}_{jk}}\,,\\
\epsilon_{UV}{\cal L}^{\phi \sigma^2}_{5PN}&=&\paq{\frac23 E\pa{\dddot{\cal T}^{ijaa}\dot{\cal T}^{kl}+\ddot{\cal T}^{ija}\ddot{\cal T}^{kla}}+\frac43 \dot{G}^a \ddot{\cal T}^{ija}\dot{\cal T}^{kl}+\frac13\ddot{Q}\pa{\dot{\cal T}^{ij}\dot{\cal T}^{kl}}}\left(\delta_{ij} \delta_{kl}-\delta_{ik} \delta_{jl}\right)\,,\nn\\
\epsilon_{UV}{\cal L}^{A^2 \sigma}_{5PN}&=&\frac8{15}\pa{\dot{P}^k \dddot{M}^{kij}+2\ddot{M}^{ki}\ddot{M}^{kj}}\left({\cal T}^{ij}-2{\cal T}^{ll}\delta_{ij}\right)+\frac4{15}\dot{\vec{P}}^2\left(\ddot{{\cal T}}^{ijij}-2\ddot{{\cal T}}^{iijj}\right)\nn\\
&&+\frac{16}{15}\dot{P}^k \ddot{M}^{ki}\left(\dot{{\cal T}}^{ijj}-2\dot{{\cal T}}^{jji}\right)\,,\\
\epsilon_{UV}{\cal L}^{\sigma^3}_{5PN}&=&\frac43{\cal T}^{jj}\left[\left(\dot{{\cal T}}^{kl}\right)^2-\left(\dot{{\cal T}}^{kk}\right)^2\right]\,,\\
\epsilon_{UV}{\cal L}^{\phi^2 A}_{5PN}&=&\frac23 \dot{E}^2 \dot{M}^{kk}+\frac2{15}\dot{E}\paq{\ddddot{O}^{jjk}  {\cal P}^k+\ddot{G}^{i} \pa{\ddot{M}^{ikk}+2\ddot{M}^{kik}}
+\dddot{Q}\dot{M}^{kk}+2\dddot{Q}^{ij}\dot{M}^{ij}}\nn\\
&&+\frac1{210}\pa{\ddot{\dddot{H}}^{iijj}\ddot{G}^k {P}^k+4\ddot{\dddot{H}}^{ijkk}\ddot{G}^i {P}^j}+\frac1{105}{P}^k\pa{\dddot{Q}\ddddot{O}^{kjj}+2\dddot{Q}^{jk}\ddddot{O}^{iij}+2\dddot{Q}^{ij}\ddddot{O}^{ijk}}\nn\\
&&+\frac1{105}\ddot{G}^i\ddot{G}^j\pa{\dddot{M}^{kkll}\delta_{ij}+2\dddot{M}^{ijll}+2\dddot{M}^{llij}}+\frac1{105}\dddot{Q}^{ab}\ddot{G}^c\ddot{M}^{ijk}\delta_{abcijk}\nn\\
&&+\frac2{105}\ddot{G}^{i}\paq{\ddddot{O}^{ijj}\dot{M}^{kk}+2\ddddot{O}^{ijk}\dot{M}^{jk}+\ddddot{O}^{jkk}\left(\dot{M}^{ij}+\dot{M}^{ji}\right)}\nn\\
&&+\frac1{210}\paq{\dot{M}^{kk}\pa{\dddot{Q}^2+2\left(\dddot{Q}^{ij}\right)^2}+4\dot{M}^{ij}\pa{\dddot{Q}^{ij}\dddot{Q}+2\dddot{Q}^{ki}\dddot{Q}^{kj}}}\,,\\
\epsilon_{UV}{\cal L}^{A^3}_{5PN}&=&-\frac25 \dot{\vec{P}}^2 \dddot{M}^{jjkk}-\frac45 \ddddot{M}^{ijkk}\dot{P}^i P^j-\frac8{15} \ddot{M}^{ij}\dot{P}^i \pa{\ddot{M}^{jkk}+2\ddot{M}^{kjk}}
-\frac8{15} \ddot{M}^{ij}\ddot{M}^{ik} \pa{\dot{M}^{ll}\delta_{jk}+2\dot{M}^{jk}}\nn\\
&&+\frac8{15} \ddot{M}^{ij}P^k \pa{\dddot{M}^{ill}\delta_{jk}+2\dddot{M}^{ijk}}-\frac8{15} \dot{M}^{ij}\dot{P}^k \pa{\dddot{M}^{kll}\delta_{ij}+2\dddot{M}^{kij}}\,,\\
\epsilon_{UV}{\cal L}^{A\sigma^2}_{5PN}&=&\frac43\dot{M}^{jj}\paq{\left(\dot{{\cal T}}^{kl}\right)^2-\left(\dot{{\cal T}}^{kk}\right)^2}+\frac83P^{k}\paq{\dot{{\cal T}}^{ij}\ddot{{\cal T}}^{ijk}-\dot{{\cal T}}^{ii}\ddot{{\cal T}}^{jjk}}\,.
\ee

By setting to zero $P^i$ and $G^i$, which are respectively the total momentum
and the centre of mass position, and using the equations of motion, one
recovers the 5PN octupole and magnetic quadrupole tail poles coefficients
($-1/189$ and $-16/45$, respectively) derived in \cite{Foffa:2019eeb},
including the fact that traces of tensors disappear in the final result.
One can also verify that the terms associated
to the angular momentum tail and to the quadrupole memory have no poles.

\subsection{UV poles and power emission coefficients}

The calculation of the previous subsection provides useful information about
the far
zone-near zone interplay and pole cancellation, but as long as one is interested
to the contribution to the binary dynamics, this result is redundant because of
the presence of terms involving derivatives of the centre of mass position and
traces of tensors which are known to cancel in the physical result.
For a more systematic study of the tail and memory UV poles, it is convenient
to follow \cite{Foffa:2019eeb} and consider the on-shell action in terms of the
totally symmetric and traceless multipoles ${\cal I}^M\,,{\cal J}^M$ in the
centre of mass frame, as well as of the total angular momentum $\vec{L}$ and of
the energy $E$:
\be
\ba{rcl}
\label{eq:mult}
\ds S_{mult}&=&\ds\frac1{\Lambda} \left\{\int {\rm d}{\tau}\left[-E
-\frac12\dot x^{\mu}L_{\alpha\beta}\omega^{\alpha\beta}_{\mu}\right.\right.\\
&&\ds\qquad\qquad+\sum_{\ell\geq 2}\pa{
\frac1{\ell!}{\cal I}^{\mu_1\dots\mu_\ell}{\cal E}_{\mu_1\mu_2;\mu_3\dots\mu_\ell}-
\frac{2\ell}{(\ell+1)!}{\cal J}^{\mu_1\dots\mu_\ell}{\cal B}_{\mu_1\mu_2;\mu_3\dots\mu_\ell}}
\bigg]\Bigg{\}}\\
&\simeq&\ds\frac 1{\Lambda} \int {\rm d}t\paq{\frac 12 E h_{00}+\frac12 \epsilon_{ijk}L^i h_{0j,k}-\frac12 {\cal I}^{ij}{\cal E}_{ij}-\frac{1}{6} {\cal I}^{ijk}{\cal E}_{ij,k}-\frac{2}3 {\cal J}^{ij}{\cal B}_{ij}+\dots}\,,
\ea
\ee
where  ${\cal E}_{ij}\equiv R_{0i0j}\simeq-\frac12\left(h_{00,ij}+\ddot{h}_{ij}-\dot{h}_{0i,j}-\dot{h}_{0j,i}\right)$ and ${\cal B}_{ij}\equiv\frac12\epsilon_{ikl}R_{0jkl}$, with
$R_{0jkl}\simeq\frac12\left(\dot{h}_{jk,l}-\dot{h}_{jl,k}+h_{0l,jk}-h_{0k,jl}\right)$, being $R^\mu_{\nu\rho\sigma}$ the standard Riemann tensor.\footnote{A word of caution is needed here: in a derivation ${\it ab\ initio}$
  of eq.(\ref{eq:mult}), there are no Levi-Civita tensors; they appear in
  magnetic-like terms $\epsilon_{ijk}h_{0j,k}$, ${\cal B}_{ij}\,,\dots$ as
  shortcuts for the antisymmetric combination $\delta^{ac}\delta^{bd}-\delta^{ad}\delta^{bc}$
  which is generated by the contraction with other Levi-Civita's present in the
  standard three-dimensional definitions of the magnetic multipoles $L^i$,
  ${\cal J}^{ij}\,,\dots$. This implies the convention that such ad-hoc inserted
  Levi-Civita's behave as 3-dimensional tensors, that is
  $\delta^{ac}\delta^{bd}-\delta^{ad}\delta^{bc}=\epsilon^{iac}\epsilon^{ibd}$;
  had we adopted the $d$-dimensional formula
  $\epsilon^{iac}\epsilon^{ibd}=\pa{d-2}\pa{\delta^{ac}\delta^{bd}-\delta^{ad}\delta^{bc}}$,
  the extra $\pa{d-2}$ factor should have been compensated by introducing an
  analogous factor in the denominator of the definitions the magnetic multipoles.}

After taking some derivatives by parts, the multipolar action can be recast
into the form (\ref{eq:multgen}) with:
\begin{eqnarray}
{\cal O}_{\phi}(t)&=&-E(t)\,,\quad {\cal O}^{i_1}_{\phi}(t)=0\,,\nonumber\\
{\cal O}^{N}_{\phi}(t)&\equiv&{\cal O}^{i_1\dots i_n}_{\phi}(t)=-{\cal I}^{i_1\dots i_n}(t)\quad{\rm for}\ n\geq2\,,\nonumber\\
{\cal O}^i_{A}(t)&=& 0\,, \quad {\cal O}^{ii_1}_{A}(t)=\epsilon_{ki {i_1}}\frac{L^{k}(t)}{2}+\frac12\dot{\cal I}^{i{i_1}}(t)\,,\nonumber\\
{\cal O}^{iN}_{A}(t)&\equiv& {\cal O}^{i i_1\dots i_n}_{A}(t)=\frac1{n+1}\dot{\cal I}^{i i_1\dots i_n}(t)+\frac{n}{n+1}\epsilon_{ki i_1}{\cal J}^{k i_2\dots i_n}(t)\quad{\rm for}\ n\geq2\,,\nonumber\\
{\cal O}^{ij}_{\sigma}(t)&=&\frac12\ddot{\cal I}^{ij}\,,\nonumber\\
{\cal O}^{ij N}_{\sigma}(t)&\equiv& {\cal O}^{ij i_1\dots i_n}_{\sigma}(t)=\frac1{(n+1)(n+2)} \ddot{\cal I}^{iji_1\dots i_n}(t)+\frac{2}{n+2}\epsilon_{i_1k(i}\dot{\cal J}^{j)ki_2\dots i_n}\quad{\rm for}\ n\geq1\,.\nonumber
\end{eqnarray}

It is useful to separate contributions coming from $E$, $\vec{L}$, and ${\cal I}^N, {\cal J}^N$;
in doing this, we report only those terms that are not vanishing after taking into account traceleness, tensor symmetries  and conservation equations.
Starting from the contributions of diagrams involving the energy $E$ and two
electric multipoles, one finds
\be
{\epsilon}_{UV}{\cal L}^{\phi^3}_{E{\cal I}^2}&=&-\ell!F^e_{2\ell}[\ell,\ell]E\left({\cal I}^{(\ell+1)}_{i_1\dots i_{\ell}}\right)^2\nonumber\\
{\epsilon}_{UV}{\cal L}^{\phi \sigma^2}_{E{\cal I}^2}&=&-\frac{2(\ell-2)!}{\ell^2(\ell-1)^2}F^e_{2\ell-4}[\ell-2,\ell-2]E\left({\cal I}^{(\ell+1)}_{i_1\dots i_{\ell}}\right)^2
=-2(2\ell+1)(2\ell-1)(\ell-2)!F^e_{2\ell}[\ell,\ell]E\left({\cal I}^{(\ell+1)}_{i_1\dots i_{\ell}}\right)^2\nonumber\\
{\epsilon}_{UV}{\cal L}^{\phi A^2}_{E{\cal I}^2}&=&\frac{4(\ell-1)!}{\ell^2}F^e_{2\ell-2}[\ell-1,\ell-1]E\left({\cal I}^{(\ell+1)}_{i_1\dots i_{\ell}}\right)^2
=4(2\ell+1)(\ell-1)!F^e_{2\ell}[\ell,\ell]E\left({\cal I}^{(\ell+1)}_{i_1\dots i_{\ell}}\right)^2\nonumber\,,
\ee
with $F^e_n$ defined in Appendix \ref{app:poles_her} (a new set of collective indices $\ell$ has been introduced for convenience, keeping track of the rank of the multipole tensors);
the sum of the three contributions gives exactly (minus) the Thorne coefficients for electric multipole radiation $c^{\cal I}_{\ell}=\frac{2\ell(\ell+1)(\ell+2)}{\ell(\ell-1)(2 \ell+1)!}$, as expected.

Analogously, for the tail involving the conserved energy and two magnetic multipoles,
\be
{\epsilon}_{UV}{\cal L}^{\phi \sigma^2}_{E{\cal J}^2}&=&-\frac{4(\ell-2)!}{\ell+1}F^e_{2\ell-2}[\ell-1,\ell-1]E\left({\cal J}^{(l)}_{i_1\dots i_{\ell}}\right)^2
=-\frac{4\ell^2(2\ell+1)(\ell-2)!}{\ell+1}F^e_{2\ell}[\ell,\ell]E\left({\cal J}^{(l)}_{i_1\dots i_{\ell}}\right)^2\nonumber\\
{\epsilon}_{UV}{\cal L}^{\phi A^2}_{E{\cal J}^2}&=&\frac{4\ell\ \ell!}{\ell+1}F^e_{2\ell}[\ell,\ell]E\left({\cal J}^{(l)}_{i_1\dots i_{\ell}}\right)^2\nonumber\,,
\ee
and the total gives again the (opposite of the) appropriate Thorne radiation coefficients  $c^{\cal J}_{\ell}=\frac{2^{\ell+3}\ell(\ell+2)}{(\ell-1)(2 \ell+2)!}$.

All terms involving the angular momentum $\vec{L}$ (and two dynamical multipoles) have vanishing divergent part after using the conservation equation, the
next case to study are memory processes, which involve only radiative multipoles.
Three electric multipole cases give
\be
{\epsilon}_{UV}{\cal L}^{\phi^3}_{{\cal I}_1{\cal I}_2{\cal I}_3}&=&-F^e_{\ell_1+\ell_2+\ell_3}[\ell_1,\ell_2]
\langle{\cal I}^{(\ell_1+1)}_{[\ell_1]}{\cal I}^{(\ell_2+1)}_{[\ell_2]}{\cal I}^{(\ell_3)}_{[\ell_3]}\rangle\nonumber\\
{\epsilon}_{UV}{\cal L}^{\phi^2\sigma}_{{\cal I}_1{\cal I}_2{\cal I}_3}&=&-\frac{F^e_{\ell_1+\ell_2+\ell_3-2}[\ell_1,\ell_2]}{\ell_1+\ell_2+\ell_3+1}
\langle{\cal I}^{(\ell_1+1)}_{[\ell_1]}{\cal I}^{(\ell_2+1)}_{[\ell_2]}\frac{{\cal I}^{(\ell_3)}_{[\ell_3]}}{(\ell_3-1)\ell_3}\rangle\nonumber\\
{\epsilon}_{UV}{\cal L}^{\phi\sigma^2}_{{\cal I}_1{\cal I}_2{\cal I}_3}&=&-2F^e_{\ell_1+\ell_2+\ell_3-4}[\ell_1-2,\ell_2-2]
\langle\frac{{\cal I}^{(\ell_1+1)}_{ij[\ell_1-2]}}{\ell_1(\ell_1-1)}\frac{{\cal I}^{(\ell_2+1)}_{ij[\ell_2-2]}}{\ell_2(\ell_2-1)}{\cal I}^{(\ell_3)}_{[\ell_3]}\rangle\nonumber\\
{\epsilon}_{UV}{\cal L}^{\phi A^2}_{{\cal I}_1{\cal I}_2{\cal I}_3}&=&4F^e_{\ell_1+\ell_2+\ell_3-2}[\ell_1-1,\ell_2-1]
\langle\frac{{\cal I}^{(\ell_1+1)}_{i[\ell_1-1]}}{\ell_1}\frac{{\cal I}^{(\ell_2+1)}_{i[\ell_2-1]}}{\ell_2}{\cal I}^{(\ell_3)}_{[\ell_3]}\rangle\nonumber\\
{\epsilon}_{UV}{\cal L}^{A^2\sigma}_{{\cal I}_1{\cal I}_2{\cal I}_3}&=&4\frac{F^e_{\ell_1+\ell_2+\ell_3-4}[\ell_1-1,\ell_2-1]}{\ell_1+\ell_2+\ell_3-1}
\langle\frac{{\cal I}^{(\ell_1+1)}_{i[\ell_1-1]}}{\ell_1}\frac{{\cal I}^{(\ell_2+1)}_{i[\ell_2-1]}}{\ell_2}\frac{{\cal I}^{(\ell_3)}_{[\ell_3]}}{\ell_3(\ell_3-1)}\rangle\nonumber\\
{\epsilon}_{UV}{\cal L}^{\sigma^3}_{{\cal I}_1{\cal I}_2{\cal I}_3}&=&-2\frac{F^e_{\ell_1+\ell_2+\ell_3-6}[\ell_1-2,\ell_2-2]}{\ell_1+\ell_2+\ell_3-3}
\langle\frac{{\cal I}^{(\ell_1+1)}_{ij[\ell_1-2]}}{\ell_1(\ell_1-1)}\frac{{\cal I}^{(\ell_2+1)}_{ij[\ell_2-2]}}{\ell_2(\ell_2-1)}\frac{{\cal I}^{(\ell_3)}_{[\ell_3]}}{\ell_3(\ell_3-1)}\rangle\nonumber\\
{\epsilon}_{UV}{\cal L}^{\phi^2 A}_{{\cal I}_1{\cal I}_2{\cal I}_3}&=&2F^o_{\ell_1+\ell_2+\ell_3-1}[\ell_1,\ell_2]
\langle{\cal I}^{(\ell_1+1)}_{[\ell_1]}{\cal I}^{(\ell_2+1)}_{[\ell_2]}\frac{{\cal I}^{(\ell_3)}_{[\ell_3]}}{\ell_3}\rangle\nonumber\\
{\epsilon}_{UV}{\cal L}^{A^3}_{{\cal I}_1{\cal I}_2{\cal I}_3}&=&-8F^o_{\ell_1+\ell_2+\ell_3-3}[\ell_1-1,\ell_2-1]
\langle\frac{{\cal I}^{(\ell_1+1)}_{i[\ell_1-1]}}{\ell_1}\frac{{\cal I}^{(\ell_2+1)}_{i[\ell_2-1]}}{\ell_2}\frac{{\cal I}^{(\ell_3)}_{[\ell_3]}}{\ell_3}\rangle\nonumber\\
{\epsilon}_{UV}{\cal L}^{A\sigma^2}_{{\cal I}_1{\cal I}_2{\cal I}_3}&=&4F^o_{\ell_1+\ell_2+\ell_3-5}[\ell_1-2,\ell_2-2]
\langle\frac{{\cal I}^{(\ell_1+1)}_{ij[\ell_1-2]}}{\ell_1(\ell_1-1)}\frac{{\cal I}^{(\ell_2+1)}_{ij[\ell_2-2]}}{\ell_2(\ell_2-1)}\frac{{\cal I}^{(\ell_3)}_{[\ell_3]}}{\ell_3}\rangle\nonumber\,,
\ee
with $F^o_n$  and $\langle\ \rangle$ defined in Appendix \ref{app:poles_her}.

The above contributions can be simplified to show that it is exactly zero.
Note that three subgroups of the sum are separately vanishing:
${\cal L}^{x^2\phi}_{{\cal I}_1{\cal I}_2{\cal I}_3}={\cal L}^{x^2\sigma}_{{\cal I}_1{\cal I}_2{\cal I}_3}=-\frac12{\cal L}^{x^2A}_{{\cal I}_1{\cal I}_2{\cal I}_3}$, being $x$
any of $\phi,A,\sigma$, and ${\cal L}_{{\cal I}_1{\cal I}_2{\cal I}_3}^{\phi A\sigma}=0$.
Indeed, by playing with the definitions of $F^e_n, F^o_n$ one can write:
\be
{\epsilon}_{UV}\pa{{\cal L}^{\phi^3}_{{\cal I}_1{\cal I}_2{\cal I}_3}+{\cal L}^{\phi^2A}_{{\cal I}_1{\cal I}_2{\cal I}_3}+{\cal L}^{\phi^2 \sigma}_{{\cal I}_1{\cal I}_2{\cal I}_3}}&=&
2C[\ell_1,\ell_2,\ell_3]\langle{\cal I}^{(\ell_1+1)}_{[\ell_1]}{\cal I}^{(\ell_2+1)}_{[\ell_2]}{\cal I}^{(\ell_3)}_{[\ell_3]}\rangle\nonumber\\
{\epsilon}_{UV}\frac{{\cal L}^{\phi A^2}_{{\cal I}_1{\cal I}_2{\cal I}_3}+{\cal L}^{A^3}_{{\cal I}_1{\cal I}_2{\cal I}_3}+{\cal L}^{A^2 \sigma}_{{\cal I}_1{\cal I}_2{\cal I}_3}}{\ell_1+\ell_2+\ell_3+1}&=&
8C[\ell_1,\ell_2,\ell_3]\langle{\cal I}^{(\ell_1+1)}_{i[\ell_1-1]}{\cal I}^{(\ell_2+1)}_{i[\ell_2-1]}{\cal I}^{(\ell_3)}_{[\ell_3]}\rangle\nonumber\\
{\epsilon}_{UV}\frac{{\cal L}^{\phi\sigma^2}_{{\cal I}_1{\cal I}_2{\cal I}_3}+{\cal L}^{A\sigma^2}_{{\cal I}_1{\cal I}_2{\cal I}_3}+{\cal L}^{\sigma^3}_{{\cal I}_1{\cal I}_2{\cal I}_3}}{(\ell_1+\ell_2+\ell_3+1)(\ell_1+\ell_2+\ell_3-1)}&=&
4C[\ell_1,\ell_2,\ell_3]\langle{\cal I}^{(\ell_1+1)}_{ij[\ell_1-2]}{\cal I}^{(\ell_2+1)}_{ij[\ell_2-2]}{\cal I}^{(\ell_3)}_{[\ell_3]}\rangle\nonumber\\
\ee
and the common factor $C[\ell_1,\ell_2,\ell_3]\equiv\left(\frac1{\ell_3}F^o_{\ell_1+\ell_2+\ell_3-1}[\ell_1,\ell_2]-F^e_{\ell_1+\ell_2+\ell_3}[\ell_1,\ell_2]\right)$ on the
right hand side is identically vanishing.

Next, the case involving one electric and two magnetic multipoles
\be
{\epsilon}_{UV}{\cal L}^{\phi\sigma^2}_{{\cal J}_1{\cal J}_2{\cal I}_3}&=&2F^e_{\ell_1+\ell_2+\ell_3-4}[\ell_1-2,\ell_2-2]
\langle4\left[\frac{{\cal J}^{(\ell_1+1)}_{i[\ell_1-1]}}{\ell_1+1}\frac{{\cal J}^{(\ell_2+1)}_{i[\ell_2-1]}}{\ell_2+1}
-\frac{{\cal J}^{(\ell_1+1)}_{ij[\ell_1-2]}}{\ell_1+1}\frac{{\cal J}^{(\ell_2+1)}_{ij[\ell_2-2]}}{\ell_2+1}\delta_{i_1 i_2}\right]{\cal I}^{(\ell_3)}_{[\ell_3]}\rangle\nonumber\\
{\epsilon}_{UV}{\cal L}^{\phi A^2}_{{\cal J}_1{\cal J}_2{\cal I}_3}&=&4F^e_{\ell_1+\ell_2+\ell_3-2}[\ell_1-1,\ell_2-1]
\langle\left[\frac{\ell_1{\cal J}^{(\ell_1+1)}_{i[\ell_1-1]}}{\ell_1+1}\frac{\ell_2{\cal J}^{(\ell_2+1)}_{i[\ell_2-1]}}{\ell_2+1}\delta_{i_1 i_2}
-\frac{\ell_1{\cal J}^{(\ell_1+1)}_{[\ell_1]}}{\ell_1+1}\frac{\ell_2{\cal J}^{(\ell_2+1)}_{[\ell_2]}}{\ell_2+1}\right]{\cal I}^{(\ell_3)}_{[\ell_3]}\rangle\nonumber\\
{\epsilon}_{UV}{\cal L}^{A^2\sigma}_{{\cal J}_1{\cal J}_2{\cal I}_3}&=&4\frac{F^e_{\ell_1+\ell_2+\ell_3-4}[\ell_1-1,\ell_2-1]}{\ell_1+\ell_2+\ell_3-1}
\langle\left[\frac{\ell_1{\cal J}^{(\ell_1+1)}_{i[\ell_1-1]}}{\ell_1+1}\frac{\ell_2{\cal J}^{(\ell_2+1)}_{i[\ell_2-1]}}{\ell_2+1}\delta_{i_1 i_2}
-\frac{\ell_1{\cal J}^{(\ell_1+1)}_{[\ell_1]}}{\ell_1+1}\frac{\ell_2{\cal J}^{(\ell_2+1)}_{[\ell_2]}}{\ell_2+1}\right]\frac{{\cal I}^{(\ell_3)}_{[\ell_3]}}{\ell_3(\ell_3-1)}\rangle\nonumber\\
{\epsilon}_{UV}{\cal L}^{\sigma^3}_{{\cal J}_1{\cal J}_2{\cal I}_3}&=&2\frac{F^e_{\ell_1+\ell_2+\ell_3-6}[\ell_1-2,\ell_2-2]}{\ell_1+\ell_2+\ell_3-3}
\langle4\left[\frac{{\cal J}^{(\ell_1+1)}_{i[\ell_1-1]}}{\ell_1+1}\frac{{\cal J}^{(\ell_2+1)}_{i[\ell_2-1]}}{\ell_2+1}
-\frac{{\cal J}^{(\ell_1+1)}_{ij[\ell_1-2]}}{\ell_1+1}\frac{{\cal J}^{(\ell_2+1)}_{ij[\ell_2-2]}}{\ell_2+1}\delta_{i_1 i_2}\right]\frac{{\cal I}^{(\ell_3)}_{[\ell_3]}}{\ell_3(\ell_3-1)}\rangle\nonumber\\
{\epsilon}_{UV}{\cal L}^{A^3}_{{\cal J}_1{\cal J}_2{\cal I}_3}&=&8F^o_{\ell_1+\ell_2+\ell_3-3}[\ell_1-1,\ell_2-1]
\langle\left[\frac{\ell_1{\cal J}^{(\ell_1+1)}_{[\ell_1]}}{\ell_1+1}\frac{\ell_2{\cal J}^{(\ell_2+1)}_{[\ell_2]}}{\ell_2+1}
-\frac{\ell_1{\cal J}^{(\ell_1+1)}_{i[\ell_1-1]}}{\ell_1+1}\frac{\ell_2{\cal J}^{(\ell_2+1)}_{i[\ell_2-1]}}{\ell_2+1}\delta_{i_1 i_2}\right]\frac{{\cal I}^{(\ell_3)}_{[\ell_3]}}{\ell_3}\rangle\nonumber\\
{\epsilon}_{UV}{\cal L}^{A\sigma^2}_{{\cal J}_1{\cal J}_2{\cal I}_3}&=&4F^o_{\ell_1+\ell_2+\ell_3-5}[\ell_1-2,\ell_2-2]
\langle4\left[\frac{{\cal J}^{(\ell_1+1)}_{ij[\ell_1-2]}}{\ell_1+1}\frac{{\cal J}^{(\ell_2+1)}_{ij[\ell_2-2]}}{\ell_2+1}\delta_{i_1 i_2}
-\frac{{\cal J}^{(\ell_1+1)}_{i[\ell_1-1]}}{\ell_1+1}\frac{{\cal J}^{(\ell_2+1)}_{i[\ell_2-1]}}{\ell_2+1}\right]\frac{{\cal I}^{(\ell_3)}_{[\ell_3]}}{\ell_3}\rangle\nonumber\,,
\ee
and again the sum of the contributions gives zero (separately in two groups)
The cases in which a magnetic multipole carry the index set $\ell_3$ gives identically vanishing contractions, as well as the cases involving an odd number of magnetic multipoles.

In conclusion, the poles of all the hereditary contributions not involving
the conserved energy $E$ vanish, in agreement with
the standard result that memory terms do not affect the luminosity
function \cite{Blanchet:2013haa,Marchand:2016vox}, even though they affect the
waveform phase.

\section{Summary and conclusions}
\label{sec:summ}

The interplay between radiative and potential modes is
 an intriguing aspects of
 two body dynamics,
its first manifestation being represented by the famous tail effect
\cite{Blanchet:1987wq}
appearing at fourth post-Newtonian order in the conservative dynamics.
As the study of such systems is rapidly pacing towards higher post-Newtonian
sectors, which are being unveiled
by the combined use of traditional General Relativity approaches and new
techniques imported from particle physics and field theory in general,
  we find useful to summarise and clarify the subtleties associated with
  the momentum region decomposition inherent to the splitting into near
  and far zone.

  We considered two-body interaction processes contributing to the conservative
  dynamics, without radiation emission at infinity.
Within this setup, our first investigation concerned the kind of boundary conditions to be used in
the Green's functions of the radiative modes, the choice being
immaterial for longitudinal ones.

Reminding that
  dynamics defines the equations of motion, but their solutions require
  a physically motivated choice of boundary conditions which is independent
  of the fundamental theory encoded in the Lagrangian, we argued that
  when radiative modes are involved one needs to use causal
  (i.e. advanced and/or retarded) Green's functions to compute the effective
  Lagrangian. We also showed that for processes involving up to two causal
  Green's function, they can actually be substituted with Feynman ones
  with the bonus of obtaining the imaginary part related to the
  energy dissipated in the process (without the needs to use closed-time-path
  formalism which however is necessary to compute radiation reaction forces).

Another key aspect of the success of the post-Newtonian program
is the use of the method of regions to simplify the integrals
needed to solve equations of motions.
In this approach the two-body problem is broken into two simpler sub-problems:
a near zone one where the two binary constituents interact with simplified Green's
functions and a far zone one in which exact Green's functions are attached to
a multipole-expanded composite source.

While the decomposition is simple for the case of no bulk interactions,
we have shown how diagrams containing one bulk triple interaction is
decomposed into a near zone contribution and a far zone one, both
contributing to the conservative dynamics, plus a purely
imaginary term corresponding to a correction to the leading order
self-interaction diagram, which does not belong to the conservative sector.
Scale separation and momentum conservation have been crucially used in the
proof, and one may wonder about the generalisation of the argument to more
complicated, higher $G$-order diagrams.
Considering for instance a diagram containing a quadruple bulk vertex, and
thus four Green's functions, the same procedure which led to
the decomposition of diagrams with a triple bulk vertex into a near$^3$,
a far$^3$ and a near$^2$-far term, see eq.~(\ref{eq:full_simple}),
would lead to near$^4$, far$^4$, near$^3$-far and near$^2$-far$^2$ terms,
the last one potentially mixing near and far zone conservative dynamics.

However, as we have seen that mixed terms of the type near$^n$-far correct the
imaginary self-energy process, in general we expect that terms near$^n$-far$^m$ with $m>1$
represent near zone corrections to the $m$ multipoles used as sources
of the far-zone Green's functions.

Finally, we have studied the properties of far zone diagrams with three sources and a cubic bulk vertex at generic post-Newtonian order.
In particular it was shown that tail diagrams involving three
far zone Green's functions, one of which sourced by a conserved multipole,
two by dynamical ones (hence describing the process of emission, scattering by static curvature
and absorption of radiation)
can be related at all post-Newtonian orders to the luminosity function associated
to the two dynamical multipoles and also to the infra-red divergences
of the near zone, thus providing a highly non-trivial
self-consistency constraint to near and far zone computations.
On the other hand we have shown at all post-Newtonian orders that the so-called memory diagrams, i.e. those involving 
three dynamical sources, contribute to conservative dynamics but not to the
dissipative one.

\section*{Acknowledgments}
S.~F. is supported by the Fonds National Suisse and by the SwissMAP NCCR National Center of Competence in Research.
R.~S. is partially supported by CNPq.
We thank J.~Blumlein, A.~Maier, P.~Marquard and G.~Schaefer for several exchanges about the importance of boundary conditions in NRGR.

\appendix
\section{Regions decomposition}
\subsection{Tree-level diagram}
\label{app:tree}

We provide here the missing steps to derive (\ref{eq:1propNF}) from (\ref{eq:1prop}), which
can be trivially rewritten as
\be
\ba{rcl}
\ds {\mathcal S}_1&=&\ds\sum_{a,b=1}^2\int_{t_a,t_b}\int_k C_aC_b\Big[
  e^{ik_0t_{ab}-i\K \cdot\X_{ab}}N(k)\\
  &&\ds+\left.\sum_{r,s\geq 0}\frac{\pa{-i\K\cdot \X_a}^r}{r!}\frac{\pa{i\K\cdot \X_b}^s}{s!}
  \frac 12\pa{\frac 1{R(k)}+\frac 1{A(k)}}+S_{a,b}(k) \right]
\ea
 \ee
where
\be
\ba{rcl}
\ds S_{a,b}(k)&\equiv&\ds e^{ik_0t_{ab}-i\K\cdot\X_{ab}}
  \paq{\frac 12\pa{\frac 1{R(k)}+\frac 1{A(k)}}-N(k)}\\
&&\ds -\sum_{r,s\geq 0}\frac{\pa{-i\K\cdot\X_a}^r}{r!}
\frac{\pa{i\K\cdot\X_b}^s}{s!}\frac 12\pa{\frac 1{R(k)}+\frac 1{A(k)}}\,,
\ea
\ee
so that deriving (\ref{eq:1propNF}) amounts to show that the integral of $S_{a,b}(k)$
vanishes.

Separating the near and far zone scale as done in Subsection \ref{ssec:hered}, i.e.
identifying a frequency scale $\kappa$ such that $r^{-1}>\kappa>\Omega$,
one can write
\be
\label{eq:1propQED}
\ba{rcl}
\ds \int_kS_{a,b}(k)&\equiv&\ds
\int_k\left\{\theta(|\K|-\kappa)\paq{e^{ik_0t_{ab}-i\K\cdot\X_{ab}}
  \pa{\frac 12\pa{\frac 1{R(k)}+\frac 1{A(k)}}-N(k)}}\right.\\
&&\ds +\theta(\kappa-|\K|)\paq{\pa{e^{ik_0t_{ab}-i\K\cdot\X_{ab}}-
\frac{\pa{-i\K\cdot\X_a}^r}{r!}\frac{\pa{i\K\cdot\X_b}^s}{s!} }
\frac 12\pa{\frac 1{R(k)}+\frac 1{A(k)}}}\\
&&\ds \left.-\frac{\pa{-i\K\cdot\X_a}^r}{r!}\frac{\pa{i\K\cdot\X_b}^s}{s!}N(k)
\right\}
\ea
\ee
where we have used that in the near zone ($|\K|>\kappa$) one can expand the Green's
functions as $N(k)$ and in the far zone the exponential factors $e^{i\K\cdot\X}$
for $|\K|<r^{-1}$, conditions which also show the vanishing of the first two lines
of eq.~(\ref{eq:1propQED}), the third line vanishing because of dimensional
regularisation.

Note that the last line of (\ref{eq:1propQED}) may contain compensating
infra-red and ultra-violet divergences.

\subsection{Hereditary terms}
\label{app:HHS}

We derive here eq.(\ref{eq:rest_props}), justifying the decomposition
of the conservative sector of the full theory (\ref{eq:3props})
into a pure near zone plus a pure far zone contribution.

Adopting the notation $\int_H\equiv\int_{\K}\theta\pa{|\K|-\kappa}$ and $\int_S\equiv\int_\K\theta\pa{\kappa-|\K|}$, one can write
\be
\int S_{a,b,c}=\paq{\int_{HHH}+\int_{SSS}} S_{a,b,c}+\paq{\int_{HHS}+\int_{HSH}+\int_{SHH}}S_{a,b,c}
\ee
where we have used that regions like $HSS$ are incompatible with scale separation and momentum conservation.
We than use that in the $HHH$ region $I_a^{(R)}I_b^{(R)}I_c^{(A)}=N_aN_bN_c$
and $F_a^{(R)}F_b^{(R)}F_c^{(A)}=D_aD_bD_c$
(for any combination of causal Green's functions), in the $SSS$
region $I_a^{(R)}I_b^{(R)}I^{(A)}_c=F_a^{(R)}F_b^{(R)}F_c^{(A)}$ and
$N_aN_bN_c=D_aD_bD_c$, and
in the $HHS$ one can replace $I_a^{(R)}I_b^{(R)}I^{(A)}_c$ with $N_aN_bF^{(R)}_c$,
$N_aN_bN_c$ with $N_aN_bD_c$ and $F_a^{(R)}F_b^{(R)}F_c^{(A)}$ with $D_aD_bF_c^{(A)}$,
to obtain
\be
\label{eq:intS3}
\ba{rcl}
S_{abc}&=&\ds-\paq{\int_{HHH}+\int_{SSS}} D_a D_b D_c+
\int_{HHS}\paq{N_a N_b\pa{F_c-D_c}-D_a D_b F_c+{\rm perms}}\\
&=&\ds-\int D_a D_b D_c+\int_{HHS}\paq{\pa{N_a N_b-D_aD_b}\pa{F_c-D_c}+{\rm perms}}
\,,
\ea
\ee
To demonstrate (\ref{eq:full_simple}), one has to preliminary observe that
\be
\ba{l}
\ds\theta\pa{|\K|-\kappa}\theta\pa{\kappa-|\pp|}\paq{N_a(k)N_b(-k-p)-D_a(k)D_b(-k-p)}V(k,-k-p,p)=\\
\ds\theta\pa{|\K|-\kappa}\theta\pa{\kappa-|\pp|}\int_{t_{a,b}}\frac{C_a C_b}{\K^2\pa{\K+\pp}^2} {\rm e}^{i k_0 t_a-i \pa{k_0+p_0}t_b}\sum_{n_a\geq 0}\pa{\frac{k_0^2}{\K^2}}^{n_a}\sum_{n_b\geq 0}\pa{\frac{\pa{k_0+p_0}^2}{\pa{\K+\pp}^2}}^{n_b}\\
\ds\times\paq{{\rm e}^{-i\K\cdot\X_a}{\rm e}^{i\pa{\K+\pp}\cdot\X_b}-\sum_{r\geq 0}\frac{\pa{-i\K.\X_a}^r}{r!}\sum_{s\geq 0}\frac{\paq{i\pa{\K+\pp}\cdot\X_b}^s}{s!}}
V(k,-k-p,p)\\
\ds=\theta\pa{|\K|-\kappa}\theta\pa{\kappa-|\pp|}\paq{N_a(k)N_b(-k)-D_a(k)D_b(-k)}\sum_{s\geq 0}\frac{\pa{i\pp\cdot\X_b}^{s}}{s!}\\
\qquad \times V((k_0,\K),(-k_0-p_0,-\K),(p_0,\pp)) P_\pp\,,
\ea
\ee
with
\be
\label{eq:p_sublead}
P_{\pp}\simeq 1+{\cal O}\pa{\frac{\pp\cdot\K}{\K^2},\frac{\pp^2}{\K^2},\frac{k_0^2}{\K^2},\frac{p_0^2}{\K^2}}\,.
\ee
Moreover, observing that $F_c^{(R)}=D_c$ in the hard region and
$N_a=D_a$ in the soft region, the integral over the $HHS$ region in
(\ref{eq:intS3}) can be extended to the entire momentum space, 
one can now finally write explicitly the
third term in equation (\ref{eq:full_simple}), i.e. $N_aN_bF_c^{(A)}$, as
\be
\label{eq:mixreg}
\ba{l}
\ds\frac 1{\Lambda^2}\int_{k,p}\int_{t_{a,b,c}}{\rm e}^{i\paq{k_0t_{ab}-p_0t_{bc}}}{\rm e}^{-i\K.\X_{ab}}\frac{C_aC_bC_c}{\K^4}
\sum_{n_a,n_b\geq 0}\pa{\frac{k_0^2}{\K^2}}^{n_a+n_b}\\
\ds\times\quad\sum_{r\geq 0}\frac{\pa{i\pp\cdot\X_b}^r}{r!}
\sum_{s\geq 0}\frac{\pa{-i\pp\cdot\X_c}^s}{s!}
\frac{P_\pp}{A(p)}V((k_0,\K),(-k_0-p_0,-\K),(p_0,0))\,,
\ea
\ee
where it has been used that integrals of $N_aN_bD_c$ and $D_aD_bD_c$ over the
entire momentum region are scaleless and hence they all vanish in dimensional
regularisation.\footnote{In GR all self-interaction vertices are
  local polynomial, so they do not alter the property of an integral
  being scale-less.}

Focusing on the leading order in $\frac{\pp\cdot\K}{\K^2}$, $\frac{p_0^2}{\K^2}$,
$\frac{k_0^2}{\K^2}$, one has
\bdm
\frac 1{\Lambda^4}\int_{k_0,p_0}\int_{t_{a,b,c}}{\rm e}^{i\paq{k_0t_{ab}-p_0t_{bc}}}C_aC_bC_c\int_{\K,\pp}
\frac{{\rm e}^{-i\K.\X_{ab}}}{\K^{4}A(p)}V((k_0,\K),(-k_0-p_0,-\K),(p_0,0))\,,
\edm
observing that integrals over $\K,\pp$ factorise and using eq.~(\ref{eq:fund}),
one has that the leading order of the integral of the last piece of
(\ref{eq:full_simple}) turns into
\be
\label{mixgen}
&&\frac 1{\Lambda^4}\int_{k_0,p_0}\pa{i p_0}\int_{t_{a,b,c}}{\rm e}^{i\paq{k_0t_{ab}-p_0t_{bc}}}C_aC_bC_c\int_{\K}
\frac{{\rm e}^{-i\K.\X_{ab}}}{\K^{4}}V((k_0,\K),(-k_0-p_0,-\K),(p_0,0))\,,
\ee
which is purely imaginary, hence it does not contribute to the conservative
dynamics.
The property of being purely imaginary is shared by all the subleading terms
contained in $P_\pp$ (\ref{eq:p_sublead}) and in the sums of
$(i\pp\cdot \X_{b,c})$ terms in expression (\ref{eq:mixreg}), completing the
demonstration that the contribution of the full integration (\ref{eq:3props})
to the conservative dynamics can be written as near-zone plus a far zone contribution.

\section{Structure of hereditary terms in Keldysh variables}
\label{app:Keldysh}

We schematically show how the causal patterns used in this work emerge from a treatment in terms of Keldysh variables \cite{Keldysh:1964ud}.

Such framework adopts a generalisation of the Hamilton's variational principle
similar to the closed-time-path formalism
(first proposed in \cite{Schwinger:1960qe}, see \cite{deWitt} for a review)
as described in \cite{Galley:2012hx},
which requires a doubling of the field variables, one propagating forward in time, the other backward, with the physical dynamics
eventually recovered by taking the difference of the two Lagrangian's of the
two fields, and identifying the two variables only after deriving the equations
of motions.
With this procedure one can obtain time-asymmetry at the level of the action, and
derive a dynamics with no time symmetry.
This procedure can also provide the time-symmetric part, which appears as the difference of the two effective Lagrangian's of the
two fields, and can thus be seen as the doubling of a conservative Lagrangian term.

For the tail process, in terms of Keldysh variables the relevant couplings are
\be
\langle\left(Q_- R^-+Q_+ R^+\right) \left(Q_- R^-+Q_+ R^+\right)M\phi V_{bulk}\rangle
\ee
with $V_{bulk}=h^+ h^- \phi$, $Q_{+}\equiv \frac12 \left[Q_1+Q_2\right]$, $Q_{-}\equiv Q_1-Q_2$ and $R^{\pm}$ standing for the $\pm$ combinations of the electric part of the Riemann tensor;
moreover since the $\phi$ source by the conserved mass is never on-shell, it
does not need the doubling procedure.
Also, $R^{\pm}\sim \partial^2 h^{\pm}$, so one is left with
\be
\langle 2(Q_- R^-) (Q_+ R^+) M \phi h^{+}h^-\phi\rangle\,.
\ee
Alternatively, the monopole coupling doubles to $M\phi_-=M\phi^+$, and in the bulk vertex one has
\be
h_1^2 \phi_1-h_2^2\phi_2=h_+^2\phi_-+2h_+h_-\phi_++\frac18h_-^2\phi_-={h^-}^2\phi^++2h^-h^+\phi^-+\frac18{h^+}^2\phi^+\,.
\ee
Since the Green's function $G^{++}=0$, the terms containing $\phi^+\phi^+$ vanishes and one is left again with
\be
\langle 2(Q_- R^-) (Q_+ R^+) M \phi^+ h^{+}h^-\phi^-\rangle\,.
\ee

For the memory term
\be
\langle (QR) (QR) (QR) V_{bulk}\rangle\rightarrow\langle  \left(Q_- R^-+Q_+ R^+\right) \left(Q_- R^-+Q_+ R^+\right) \left(Q_- R^-+Q_+ R^+\right) V_{bulk}\rangle\nonumber\,,
\ee
with $V_{bulk}=h_1^3-h_2^3$.
The only terms which survives the Wick contraction with the same constraints as above ($G^{++}=0$ and only one $Q_-$) is
\be
3\langle (Q_- h^-)(Q_+ h^+)(Q_+ h^+) h^- h^- h^+ \rangle\,,
\ee
which gives  one retarded $(G^{-+})$ and two advanced $(G^{+-})$ Green's functions.
The factor of $2(3)$ in front of the tail(memory) contributions are there to
compensate for the fact that the equations of motion are obtained by taking
the variation with respect to $Q_-$ only, and then $Q_-$ set to 0 so only terms
linear in $Q_-$ in the Lagrangian matter.

\section{Useful integrals}
\label{app:mi}
We report here the standard 1-loop master integral used in Subsection \ref{ssec:phicube} is
\be
\label{eq:Im}
\int_\K\frac 1{\pa{\pp+\K}^{2a}\K^{2b}}=\frac{\pa{\pp^2}^{d/2-a-b}}{\pa{4\pi}^{d/2}}
\frac{\Gamma(d/2-a)\Gamma(d/2-b)\Gamma(a+b-d/2)}{\Gamma(a)\Gamma(b)\Gamma(d-a-b)}\,,
\ee
which has IR divergence for $a\to d/2$ or $b\to d/2$, and UV divergence for
$a+b\to d/2$.
We also used the Fourier integral
\be
\label{eq:masterF}
\int_\K\frac{{\rm e}^{i\K\cdot \X}}{\K^{2a}}=2^{-2a}\pi^{-d/2}
\frac{\Gamma(d/2-a)}{\Gamma(a)}r^{2a-d}\,,
\ee
with $r\equiv |\X|$, which has IR divergence only, for $a\to d/2$.
Scale-less integrals vanish in dimensional regularisation, but for the sake
of tracking IR and UV divergences in (\ref{eq:ir_uv}), one
needs
\be
\int_\K\frac  1{\pa{\K^2}^{3-d/2}}\to\frac 1{4\pi^2}\pa{
  \frac 1{\epsilon_{IR}}-\frac 1{\epsilon_{UV}}}\,.
\ee

Double integrals involving three Green's functions are required by calculation
of memory-type processes like (\ref{eq:3props}).
By defining
\be
\frac 1{{\cal D}_{\alpha,\beta,\gamma}}\equiv\paq{\pa{\Q^2-q_0^2}^\alpha\pa{(\K+\Q)^2-p_0^2}^\beta\pa{\K^2-k_0^2}^\gamma}^{-1}\,,
\ee
(the case $p_0=-(k_0+q_0)$ will be needed, but it is instructive to keep it generic) and ${\cal D}\equiv{\cal D}_{1,1,1}$, the master integral needed in this
case is
\be
\label{eq:Imem}
I_m\equiv \int_{\K,\Q}\frac1{\cal D}\equiv
\int_{\K,\Q}\frac 1{\Q^2-q_0^2}\frac 1{(\K+\Q)^2-p_0^2}\frac 1{\K^2-k_0^2}\simeq
-\frac 1{32\pi^2\epsilon_{UV}}+O(\eps^0)\,,
\ee
where the choice of Green's function boundary conditions has been purposefully
left undefined, as the UV-divergent part of (\ref{eq:Im}) does not depend on
boundary conditions, whereas the finite part does.

Such finite part, as well the other master integral needed at $O(G_N^2)$,
depends on quantities as $\langle k_0\rangle\equiv\sqrt{-k_0^2}$,
which are dependent on the prescription adopted to displace the Green's function
pole from the real $k_0$ axis, i.e. on the Green's function boundary conditions.
Feynman, advanced and retarded boundary conditions give respectively
\be
\label{eq:Feyn_lim}
\sqrt{-\pa{k_0^2 + i {\tt a}}}&\rightarrow& -i|k_0|\,,\\
\label{eq:ret_adv}
\sqrt{-\pa{k_0^2\pm i {\tt a}k_0}^2}&\rightarrow& \mp ik_0\,.
\ee

Setting $\ds\Theta^{1/2}\equiv\frac{\Gamma\left(1-\frac d2\right)}{(4 \pi)^{d/2}}$,
one can then find
\be
\label{eq:fund}
\ds \int_\K\frac 1{\K^2-k_0^2}=\langle k_0\rangle\Theta^{1/2}\,,
\ee
which is valid for any $i{\tt a}$ prescription understood for $\langle k_0\rangle$.
Integral in (\ref{eq:fund}) is actually the only master integral required to determine the result
(\ref{eq:mem}), since the other one needed in intermediate passages
(\ref{eq:QQQret}), i.e. $I_m$ defined in (\ref{eq:Im}),
cancels when summing all polarisations in (\ref{eq:QQQret}) to recover the full
result (\ref{eq:mem}).

In terms of (\ref{eq:Im}) and (\ref{eq:fund}) one can write some useful
intermediate results
\be
\label{eq:int0}
\ba{rcl}
\ds\intkq \frac1{{\cal D}_{1,0,1}}&=&\ds\Theta \langle k_0\rangle\langle q_0\rangle\,,\\
\ds\intkq \frac1{{\cal D}_{-n,0,\alpha}}&=&\ds\intkq \frac1{{\cal D}_{\alpha,0,-n}}=0 \quad {\rm for}\quad n\geq 0\,,\\
\ds\intkq \frac1{{\cal D}_{1,-1,1}}&=&\ds(q_0^2+k_0^2-p_0^2)\intkq \frac1{{\cal D}_{1,0,1}}\,,\\
\ds\intkq \frac1{{\cal D}_{1,-2,1}}&=&\ds\paq{\pa{q_0^2+k_0^2}^2-p_0^4+\frac4d k_0^2 q_0^2}\intkq \frac1{{\cal D}_{1,0,1}}-2 p_0^2\intkq \frac1{{\cal D}_{1,-1,1}}\,,
\ea
\ee
as well as resolve the two indices sector:
\be\label{eq:2ind}
\ba{rcl}
\ds\int_\K\frac{k^ik^j}{\K^2-k_0^2}&=&\ds\frac{\delta^{ij}k_0^2}d
\sqrt{\Theta}\langle k_0\rangle\,,\\
\ds\int_{\K,\Q}k^iq^j\frac1{\cal D}&\simeq&\ds\frac{\delta^{ij}}{2d}\left\{I_m\pa{p_0^2-k_0^2-q_0^2}+\Theta\pa{\langle k_0\rangle\langle q_0\rangle-\langle p_0\rangle\langle k_0\rangle
-\langle p_0\rangle\langle q_0\rangle}\right\}\,,\\
\ds\int_{\K,\Q}k^ik^j\frac1{\cal D}&\simeq&\ds\frac{\delta^{ij}}d\left\{ I_m k_0^2+\Theta \langle p_0\rangle\langle q_0\rangle\right\}\,,\\
\ds\int_{\K,\Q}q^iq^j\frac1{\cal D}&\simeq&\ds\frac{\delta^{ij}}d\left\{ I_m q_0^2+\Theta \langle p_0\rangle\langle k_0\rangle\right\}\,.
\ea
\ee
The procedure can be generalised to incorporate more free indices, as reported
in \cite{Foffa:2019eeb}.
The results reported in the Appendix of \cite{Foffa:2019eeb} are recovered for
$\langle k_0\rangle=- i k_0\,,\langle q_0\rangle=- i q_0\,,\langle p_0\rangle=i p_0=- i (k_0+q_0)$,
that is retarded $k$ and $q$-Green's functions and advanced for $p$, whose value
is fixed by energy-momentum conservation.

\section{Poles in hereditary contributions}
\label{app:poles_her}

This Appendix contain useful formulae to compute the far zone UV poles at any desired PN order, at next-to-leading order in Newton constant.
The divergent part of the integrals involved in the calculation is independent of the boundary conditions, and reads
\begin{eqnarray}\label{genericpole}
I_{n_k,n_q}\equiv(pole\ of\ ) \int_{\K\,,\Q} k^{i_1}\dots k^{i_{n_k}} q^{i_1}\dots q^{i_{n_q}}\frac1{\cal D}=\frac1{(n_k+n_q+1)!!}k_0^{n_k} q_0^{n_q} I_m\delta_{i_{1}\dots i_{n_k+n_q}}\,,
\end{eqnarray}
with ${\cal D}$ being defined in Section.~\ref{sec:appl}.

For our convenience, we calculate the pole of a generic hereditary term
separately for each polarisation channel. We detail here the calculation for the $\phi^3$ case, and simply report the result for the other polarisation
combinations:
\be
   {\cal S}_{\phi^3far}&=&
   \ds\frac{128 \pi^2}{c_d} G^2 \intkq\frac{1}{\cal D}\sum_{\substack{n\ even\\ n_k+n_q\leq n}}(-1)^{m+1}\nonumber\\
&&  \times (i)^{n+2} \frac {k_0 q_0 (k_{i_1}\dots k_{i_{n_k}})(q_{j_1}\dots q_{j_{n_q}})[(k+q)_{p_1}\dots (k+q)_{p_m}]}{n_k!n_q!m!}{\cal O}^{i_1\dots i_{n_k}}_{\phi}{\cal O}^{j_1\dots j_{n_q}}_{\phi}{\cal O}^{p_1\dots p_{m}}_\phi\nonumber\\
   &\simeq&
   \ds 32 (\pi G)^2 I_0\sum_{\substack{n\ even\\ n_k+n_q\leq n}}\frac{(-1)^{m+1} (i)^{n+2} k_0^{n_k+1}q_0^{n_q+1}(k_0+q_0)^m}{n_k!n_q!m!(n_k+n_q+m+1)!!}\langle{\cal O}_{\phi}(k_0){\cal O}_{\phi}(q_0){\cal O}_{\phi}(-k_0-q_0)\rangle\nonumber\\
&\simeq&\ds\frac{G^2}{\epsilon}\sum_{\substack{n\ even\\ n_k+n_q\leq n}} F^e_n(n_k, n_q)\int {\rm d}t\langle{\cal O}^{(n_k+1)}_{\phi}{\cal O}^{(n_q+1)}_{\phi}{\cal O}^{(n-n_k-n_q)}_{\phi}\rangle+{O(\epsilon^0)}
\ee
here $m$ above stands for $n-n_k-n_q$ and 
\be
F^e_n(n_k, n_q)\equiv \frac1{n_k!n_q!(n-n_k-n_q)!(n+1)!!}\,,
\ee
${\cal O}$ has been defined in eq.~(\ref{eq:OpAs})
and $\langle{\cal O}_{\phi}{\cal O}_{\phi}{\cal O}_{\phi}\rangle$ is a shorthand
for the object obtained by contraction of all Lorentz indices with a totally symmetric combination
of delta functions: ${\cal O}^{i_1\dots i_{n_k}}_{\phi}{\cal O}^{i_{n_k+1}\dots i_{n_k+n_q}}_{\phi}{\cal O}^{i_{n_k+n_q+1}\dots i_n}_{\phi}\delta_{i_1\dots i_n}$.

From the calculation above one can deduce  a straightforward substitution rule
to obtain poles from amplitudes: gradients can be replaced by (minus) time
derivatives, with the appropriate contraction and combinatoric factors.

The other polarisation channels give
\be
\ba{rcl}
\ds{\cal S}_{\phi ^2 \sigma}&\simeq&\ds\frac{G^2}{\epsilon}\sum_{\substack{n\ even\\ n_k+n_q\leq n}}F^e_n(n_k, n_q)\\\
&&\ds\times\int {\rm d}t\left(\langle{\cal O}^{(n_k+1)}_{\phi}{\cal O}^{(n_q+1)}_{\phi}{\cal O}^{jj(n-n_k-n_q)}_{\sigma}\rangle-\frac{1}{n+3}\langle{\cal O}^{(n_k+1)}_{\phi}{\cal O}^{(n_q+1)}_{\phi}{\cal O}^{ij(n-n_k-n_q)}_{\sigma}\rangle_{ij}\right)
\ea
\ee
with $\langle{\cal O}_{\phi}{\cal O}_{\phi}{\cal O}^{jj}_{\sigma}\rangle\equiv{\cal O}^{i_1\dots i_{n_k}}_{\phi}{\cal O}^{i_{n_k+1}\dots i_{n_k+n_q}}_{\phi}{\cal O}^{jj i_{n_k+n_q+1}\dots i_n}_{\sigma}\delta_{i_1\dots i_n}$ and
$\langle{\cal O}_{\phi}{\cal O}_{\phi}{\cal O}^{ij}_{\sigma}\rangle_{ij}\equiv{\cal O}^{i_1\dots i_{n_k}}_{\phi}{\cal O}^{i_{n_k+1}\dots i_{n_k+n_q}}_{\phi}{\cal O}^{iji_{n_k+n_q+1}\dots i_n}_{\sigma}\delta_{ij i_1\dots i_n}$.
\be
{\cal S}_{\phi \sigma^2}&\simeq& \ds\frac{2G^2}{\epsilon}\sum_{\substack{n\ even\\ n_k+n_q\leq n}}F^e_n(n_k, n_q)\int {\rm d}t\langle{\cal O}^{ij(n_k+1)}_{\sigma}{\cal O}^{kl(n_q+1)}_{\sigma}{\cal O}^{(n-n_k-n_q)}_{\phi}\rangle\left(\delta_{ik}\delta_{jl}-\delta_{ij}\delta_{kl}\right)\,,\\
{\cal S}_{\phi A^2}&\simeq&-\ds\frac{4G^2}{\epsilon}\sum_{\substack{n\ even\\ n_k+n_q\leq n}}F^e_n(n_k, n_q)\int {\rm d}t\langle{\cal O}^{i(n_k+1)}_{A}{\cal O}^{i(n_q+1)}_{A}{\cal O}^{(n-n_k-n_q)}_{\phi}\rangle
\ee
with $\langle{\cal O}^i_{A}{\cal O}^i_{A}{\cal O}_{\phi}\rangle\equiv{\cal O}^{ii_1\dots i_{n_k}}_{A}{\cal O}^{i i_{n_k+1}\dots i_{n_k+n_q}}_{A}{\cal O}^{i_{n_k+n_q+1}\dots i_n}_{\phi}\delta_{aai_1\dots i_n}$,
\be
\ds{\cal S}_{A^2 \sigma}&\simeq&\ds-\frac{4G^2}{\epsilon}\sum_{\substack{n\ even\\ n_k+n_q\leq n}}F^e_n(n_k, n_q)\nn\\
&&\quad\times\int {\rm d}t\left\{\langle{\cal O}^{k(n_k+1)}_{A}{\cal O}^{k(n_q+1)}_{A}{\cal O}^{jj(n-n_k-n_q)}_{\sigma}\rangle
-\frac1{n+3}\langle{\cal O}^{k(n_k+1)}_{A}{\cal O}^{k(n_q+1)}_{A}{\cal O}^{ij(n-n_k-n_q)}_{\sigma}\rangle_{ij}\right\}\,,\\
\ds{\cal S}_{\sigma^3}&\simeq&\ds\frac{2G^2}{\epsilon}\!\!\!\sum_{\substack{n\ even\\ n_k+n_q\leq n}}F^e_n(n_k, n_q)
\!\!\!\int {\rm d}t\left\{\langle{\cal O}^{kl(n_k+1)}_{\sigma}{\cal O}^{kl(n_q+1)}_{\sigma}{\cal O}^{jj(n-n_k-n_q)}_{\sigma}\rangle-\langle{\cal O}^{kk(n_k+1)}_{\sigma}{\cal O}^{ll(n_q+1)}_{\sigma}{\cal O}^{jj(n-n_k-n_q)}_{\sigma}\rangle\right.\nn\\
&&\ds\quad\quad+\left.\frac1{n+3}\left[\langle{\cal O}^{kk(n_k+1)}_{\sigma}{\cal O}^{ll(n_q+1)}_{\sigma}{\cal O}^{ij(n-n_k-n_q)}_{\sigma}\rangle_{ij}-
\langle{\cal O}^{kl(n_k+1)}_{\sigma}{\cal O}^{kl(n_q+1)}_{\sigma}{\cal O}^{ij(n-n_k-n_q)}_{\sigma}\rangle_{ij}\right]\right\}\,,\\
\ds{\cal S}_{\phi ^2 A}
&\simeq&\ds\frac{2G^2}{\epsilon} \sum_{\substack{n\ odd\\ n_k+n_q\leq n}}F^o_n(n_k, n_q)\int {\rm d}t\langle{\cal O}^{(n_k+1)}_{\phi}{\cal O}^{(n_q+1)}_{\phi}{\cal O}^{j(n-n_k-n_q)}_{A}\rangle_j\,,
\ee
with
\be
F^o_n(n_k, n_q)\equiv \frac1{n_k!n_q!(n-n_k-n_q)!(n+2)!!}\,,
\ee
and $\langle{\cal O}_{\phi}{\cal O}_{\phi}{\cal O}^j_{A}\rangle_j\equiv{\cal O}^{1\dots i_{n_k}}_{\phi}{\cal O}^{i_{n_k+1}\dots i_{n_k+n_q}}_{\phi}{\cal O}^{ji_{n_k+n_q+1}\dots i_n}_{A}\delta_{ji_1\dots i_n}$,
\be
{\cal S}_{\phi A\sigma}&\simeq&
0\,,\\
{\cal S}_{A^3}&\simeq& -\frac{8 G^2}{\epsilon}\sum_{\substack{n\ odd\\ n_k+n_q\leq n}}F^o_n(n_k, n_q)\int {\rm d}t\langle{\cal O}^{j(n_k+1)}_{A}{\cal O}^{j(n_q+1)}_{A}{\cal O}^{k(n-n_k-n_q)}_{A}\rangle_k\,,\\
{\cal S}_{A\sigma^2}
&\simeq& -\frac{4 G^2}{\epsilon}\!\!\!\!\!\!\!\sum_{\substack{n\ odd\\ n_k+n_q\leq n}}\!\!F^o_n(n_k, n_q)\!\!\!\int {\rm d}t\!\left[\langle{\cal O}^{ii(n_k+1)}_{\sigma}{\cal O}^{jj(n_q+1)}_{\sigma}{\cal O}^{k(n-n_k-n_q)}_{A}\rangle_k\!-\!\langle{\cal O}^{ij(n_k+1)}_{\sigma}{\cal O}^{ij(n_q+1)}_{\sigma}{\cal O}^{k(n-n_k-n_q)}_{A}\rangle_k\!\right]\,.\nn\\
\ee

\end{document}